%% file: main_fkrb_management_science.tex
\documentclass[12pt]{article}
\usepackage[utf8]{inputenc}
\usepackage{amsmath,setspace,geometry}
\usepackage{amsthm}
\usepackage{amsfonts}
\usepackage[shortlabels]{enumitem}
\usepackage{rotating}
\usepackage{pdflscape}
\usepackage{graphicx}
\usepackage{bbm}
\usepackage{comment}
\usepackage[dvipsnames]{xcolor}
\usepackage{hyperref}
\hypersetup{colorlinks=true, linkcolor= BrickRed, citecolor = BrickRed, filecolor = BrickRed, urlcolor = BrickRed, hypertexnames = true}
\usepackage[]{natbib} 
\bibpunct[:]{(}{)}{,}{a}{}{,}
\usepackage[english]{babel}
\usepackage{float}
\usepackage{caption}
\usepackage{subcaption}
\usepackage{booktabs}
\usepackage{pdfpages}
\usepackage{threeparttable}
\usepackage{lscape}
\usepackage{bm}
\usepackage{setspace}
\title{The Option Value of Contract Duration:\\ Evidence from the U.S. Timber Market}
\author{Shosuke Noguchi\thanks{School of Commerce, Waseda University. Email: \href{mailto:shosuke.noguchi@waseda.jp}{shosuke.noguchi@waseda.jp}} \and Suguru Otani\thanks{Market Design Center, Department of Economics, University of Tokyo. Email: \href{mailto:suguru.otani@e.u-tokyo.ac.jp}{suguru.otani@e.u-tokyo.ac.jp}. This work was supported by JST ERATO Grant Number JPMJER2301 and JSPS KAKENHI 24K22604, Japan. We thank Masakazu Ishihara, Hajime Katayama, Donghyuk Kim, Yunmi Kong, Harry Paarsch, Amit Pazgal, Isabelle Perrigne, Xun Tang, and Weiqing Zhang for their valuable comments. We also thank seminar and conference participants at the Econometric Society World Congress 2025. This paper was previously circulated as ``Consumption Periods in Advance Selling Auctions: Evidence from the U.S. Timber Market''.
}}

\begin{document}

\maketitle
\begin{center}
First draft: \date{December 28, 2024}\\
\end{center}
\begin{abstract}
    \noindent {This study quantifies how contract duration influences buyers' willingness-to-pay (WTP) when they hold real options that allow them to flexibly time consumption in response to changing market conditions. Using contract data from the US timber industry, we show that buyers delay consumption to manage payoff risk. This behavior generates heterogeneous WTP across buyers. We use structural estimation to uncover the key parameters underlying the incentive to delay consumption. Using these estimates, we conduct counterfactual simulations to measure how longer contract durations shift WTP and to clarify the boundary conditions linked to project size, buyer composition, and market trends. The counterfactual simulations reveal that extending contract duration from 3 to 4 years raises seller revenue by 9-13\%, with effects amplified for larger projects and high-type buyers during the upward market trend.}
    \vspace{3mm}
    
\end{abstract}
\section{Introduction}\label{sec:introduction}
Real options embedded in business-to-business contracts are prevalent across markets for licenses, natural-resource rights, real estate development, and carbon emission allowances. Such contractual real options allow buyers to time their action within a specified period to maximize payoffs as market conditions change. Consider a manufacturer with an equipment lease that allows the firm to vary when and how intensively it uses the equipment over the lease term. Longer contract durations therefore make the buyer’s option more valuable, but for sellers they can lengthen the sales cycle and increase the opportunity cost of capital tied up in each deal.

These trade-offs make it essential for sellers to understand how contract duration shapes willingness-to-pay (WTP) and to identify the relevant boundary conditions. We address this question by developing a structural estimation framework that we apply to data on timber harvests and auctions. The timber market provides a particularly suitable setting to examine these issues, for three reasons. First, timber sales exhibit a real-option structure because tract owners sell the right to harvest within a fixed period, and we document that buyers strategically time removals in response to developments in lumber prices. Second, the data provide both detailed removal histories and auction outcomes, allowing us to link dynamic harvesting incentives directly to WTP. Third, a structural approach allows us to deal with the fact that auction values are endogenous to contract duration: sellers choose contract terms based on tract size and other unobserved characteristics, so we use tract-level counterfactual simulations to isolate the causal effect of duration on WTP.

We begin with model-free and reduced-form evidence. Harvesting decisions respond strongly to lumber prices and to the time remaining on the lease: cutting accelerates when prices spike and slows when prices soften or when expiration is far off. These dynamic incentives carry over to auction outcomes. Longer contract durations raise bidders' valuations because bidders can delay harvesting until prices improve. Different types of buyers also display distinct cutting and bidding patterns, implying that contract duration and auction design change the distribution of valuations in different ways for different buyers.

These key data patterns motivate our structural model, which combines auction bidding with post-auction behavior. Our model consists of two stages. In the first stage, we model oral auctions. Bidders are one of two types: loggers or sawmills, i.e., manufacturers. The two groups differ in how often they participate in auctions and what values they place on timber. In the second stage, we model the timber cutting, which follows the oral auction, as a real-options problem, where the flexibility to choose when and how much timber to harvest matters in the bidding strategy. For this stage, we construct a finite-period, dynamic, single-agent model in which the winning bidder chooses when and how much timber to cut in response to realized timber prices. The winning bidder prefers higher timber prices but must remove all timber by the deadline. These incentives characterize the observed cutting patterns.

We estimate the model using data on both timber auctions and timber harvesting. First, we use the variation in timber prices and the amounts of timber harvested to estimate the parameters underlying the dynamic cutting behavior, following the approach of \cite{rust1987optimal}. Second, we estimate the relative shares of loggers and sawmills leveraging the observed number of auction participants, using a simple estimation method similar to \cite{kong2020not} that does not assume a specific entry process. Third, using the estimated dynamic model and entry parameters, we recover the distribution of bidders' values based on the bidding equilibrium. Finally, we assess and validate our approach with Monte Carlo simulations.

Our estimation results reveal key differences between loggers and sawmills and highlight the importance of bidder heterogeneity in timber auctions. At the cutting stage, loggers show greater sensitivity to lumber prices, reflecting their position in the supply chain. In the bidding stage, sawmills tend to have higher valuations than loggers, implying that they are the high-type, although they exhibit more dispersed valuations within their type. As a robustness check, we estimate a random-coefficient model following \cite{fox2011simple-bef} that allows for random coefficients without dividing market participants into observed categories such as loggers and sawmills. The recovered distribution of heterogeneity displays a bimodal pattern, indicating that the observed types effectively capture the salient dimensions of variation. Together, these findings underscore the role of bidder heterogeneity in auction outcomes, providing the foundation for the counterfactual analysis of different contract durations. 

We then use the estimated model primitives to perform the simulations under different contract durations. Our counterfactual analysis demonstrates that extending the contract duration significantly increases sellers' revenue by enhancing bidders' valuations, particularly through the flexibility to adjust harvesting decisions based on future market conditions. This effect is more pronounced for stronger bidders, such as sawmills, whose valuations are particularly sensitive to longer horizons. Additionally, larger tracts magnify the revenue benefits of a longer contract duration due to the higher strategic value of deferring harvests. These findings highlight the critical role of contract length in shaping WTP. The magnitude of this effect depends on boundary conditions, including market trends, buyer type, and project size.

The managerial insights from this paper extend to business-to-consumer industries that offer flexible consumption time decisions, such as ticket sales, leasing items, and customer reward programs. A longer consumption period provides a real option: the flexibility to use the product when desired. This option becomes particularly valuable when future needs are uncertain. For instance, consider a customer reward program such as an airline mileage or loyalty point system with an expiration policy. The validity period functions as a financial option: it grants the consumer the right, but not the obligation, to redeem points for a reward within a fixed window. Just as loggers and sawmills in our setting delay harvesting to capture high lumber prices, consumers strategically delay redemption to maximize utility—waiting for seat availability on a high-value route or a specific future occasion. Extending the expiration term increases the program’s perceived value by expanding this option space, thereby enhancing the consumer's willingness to engage or pay. However, this flexibility also alters the firm’s liability structure and redemption timing. Our structural framework offers a rigorous approach to quantifying the beneficial side of this trade-off, guiding managers on how to optimize expiration terms to maximize program revenue while accounting for the heterogeneity in how consumers value and exercise the option to wait.

The remainder of the paper proceeds as follows. We close this section with a review of related literature. Section \ref{sec:data} describes the data and institutional background, Section \ref{sec:model} presents the model, Section \ref{sec:estimation} details the estimation strategy, and Section \ref{sec:estimation_results} reports the results. Section \ref{sec:counterfactual} outlines the counterfactual simulations, and Section \ref{sec:conclusion} concludes.

\paragraph{Related literature}
This paper connects to several strands of work on contract design, real options, advance selling, and auctions. Contract design has long been central to economics and management, and interest has grown recently within industrial marketing (e.g., \citealp{seshadri2004relationship,yang2017improving,petersen2018reconciling,hajdini2019contractual,feng2022contract,mols2022sign,qu2023dynamic}). These studies investigate pricing, renegotiation, and contract duration; see \cite{mouzas2013contract} and \cite{crosno2021effectiveness} for comprehensive surveys. Our focus is on contract duration as a design lever.

Seminal contributions by \cite{joskow1987contract} and \cite{crocker1991pretia} highlight two fundamental determinants of contract duration. First, relationship-specific investments favor longer contracts because repeated bargaining exposes parties to hold-up risk, as confirmed empirically by \cite{lopez2010indefinite} and \cite{gorovaia2018choice} in electronics subcontracting and franchising, respectively. Second, uncertainty about future payments pushes parties toward shorter contracts so they can adjust prices as conditions evolve; see \cite{feng2019series} for evidence that environmental uncertainty shapes both duration and price. These mechanisms reflect the real-option nature of contracts. We extend this literature by quantifying how contract duration affects transaction value perceived by buyers within a structural framework that explicitly models uncertainty. \cite{mackay2022contract} also structurally estimates the implications of contract duration in procurement auctions. However, while he focuses on the costs of duration arising from allocative inefficiency, our paper differs by quantifying the benefits of duration—specifically the real-option value perceived by buyers.

Because contract duration governs the real-option value available to buyers, our work is closely related to the broader real-options literature (see \citealp{broadie2004anniversary} for a survey). \cite{bollen1999real} embeds the product life cycle into option pricing, and our contract expiration mirrors the extreme case in which demand falls to zero. \cite{cong2020timing} shows that altering the exercise horizon changes option values and that sellers can partially influence exercise timing. We incorporate these insights into a structural model and estimate its parameters using our data. Recent structural approaches to real options include \cite{collard2013demand}, \cite{kellogg2014effect}, \cite{agerton2020learning}, and \cite{herrnstadt2024drilling}. \cite{herrnstadt2024drilling} is especially closely related to our study: both papers document that consumption clusters near contract expiration and examine how contract duration affects post-purchase actions and seller revenue. Our contribution is to link these dynamic behaviors to WTP using observed bids, data that \cite{herrnstadt2024drilling} do not observe.

Because purchases precede consumption and sellers receive no contingent payment afterward, timber contracts resemble advance-selling arrangements. The advance-selling literature dates back to \cite{shugan2000advance}; see \cite{xie2009advance} for a survey and \cite{yu2015rationing}, \cite{cachon2017advance}, and \cite{seo2020pricing} for more recent contributions. Although this literature recognizes the importance of the consumption window (e.g., \citealp{shugan2000advance}; \citealp{xie2001electronic}), it has devoted little attention to its length, which is central in our setting.

Our paper also relates to the broader empirical-auction literature that extracts economic insights from bidding data; see Chapter 2 of \cite{hortaccsu2021empirical} for a survey.
We are closest to \cite{bajari2014bidding}, \cite{ordin2019investment}, \cite{bhattacharya2022bidding}, and \cite{kong2022multidimensional}, who link auction outcomes to post-award behavior. Whereas those papers emphasize other contract features or investment choices, we focus on contract duration as a revenue-enhancing lever.

\section{Data}\label{sec:data}
We extract data from timber sales organized by the Bureau of Land Management (BLM), U.S. Department of the Interior, which is responsible for managing federal forests. In this section, we describe the timber sales process, our data, and key empirical findings.

\subsection{Timber sales process}
BLM manages forests in several states, including Oregon, Montana, Indiana, Colorado, Wyoming, California, and Utah. When growth reaches a harvestable stage, BLM auctions the right to cut the timber. Depending on the tract, BLM uses bilateral negotiation, oral auctions, or first-price sealed-bid auctions. Payment rules also vary: in scale sales bidders submit unit prices and pay the bid times the realized volume, whereas in lump-sum sales bidders submit a total price that is due regardless of actual harvest. We focus on lump-sum auctions because they dominate the data and simplify the modeling of payments.

Before an auction, BLM releases a prospectus describing the estimated species-specific volumes, reserve prices, auction rules, lease terms (duration, down payment, and so on), and tract characteristics such as location and road quality. Interested bidders can inspect the tract and prepare their bids accordingly.

Participants in the auctions are typically local loggers or sawmill manufacturers. Loggers and sawmill manufacturers have different positions in the wood supply chain. Loggers mainly cut timber and transport it to processing facilities called sawmills, where sawmill manufacturers process the transported wood into lumber or veneers. Lumber and veneers are primarily purchased by home-building companies as raw materials. In short, loggers are upstream, and sawmill manufacturers are downstream in the supply chain. 

After the auctions, BLM grants the winning bidder the right to remove timber from the tract. The contract is time-limited, implying that the winning bidder must cut all timber by the deadline. The lease duration is typically 36 months. If the winning bidder faces events such as weather, fire closure, and other related conditions that interrupt operating time, BLM may extend the lease duration. However, it is hard to accurately predict the likelihood of these factors in the future, the winning bidder must prepare to remove all timber by the due date at the time of sale.

\subsection{Data description}\label{subsec:data_description}
This paper relies on data from timber auctions conducted between 2012Q2 and 2023Q1 by the Bureau of Land Management (BLM), U.S. Department of the Interior. BLM data consist of two types of records: transaction data and contract data. The former include auction results such as the number of bidders, winning prices, reserve prices, and (when available) second-highest bids. These data also include other auction characteristics, including the number of acres of timber tract and lease duration. The latter data contain records of timber harvests, such as the timing and the volume of timber removal. We also draw data on lumber market prices from Federal Reserve Economic Data (FRED). As we see in the later section, lumber market prices relate to timber removal.

Table \ref{tb:summary_statistics_of_entry_and_bidding_stage_oral} reports summary statistics for BLM oral auctions, which constitute our primary focus. BLM sells timber using two auction formats—ascending (oral) and first-price sealed bid—as well as bilateral negotiations for small tracts. Among auctioned sales, oral auctions account for 73\% (306/418), and sealed bids account for 27\% (112/418). As documented in Appendix \ref{sec:data_appendix}, sealed‐bid contracts are systematically smaller and more geographically dispersed, implying a limited contribution to aggregate revenues relative to oral auctions; by contrast, oral auctions allocate substantially larger tracts on average. Participant feedback reportedly favors oral auctions, though they were partially suspended during the COVID-19 period, and negotiations are typically reserved for very small tracts with partners chosen on a rotational basis. Taken together, these facts motivate concentrating the main analysis on oral auctions, while providing sealed‐bid evidence in the appendix for completeness.

\paragraph{Entry and bidding stage.}
Panel (a) in Table \ref{tb:summary_statistics_of_entry_and_bidding_stage_oral} summarizes participation, prices, and tract characteristics for oral auctions (2012Q2–2023Q1). On average, about $2.16$ bidders enter per sale, while roughly $10.3$ are counted as potential participants. Because the raw files do not report potential bidders, we define this as the number of unique bidders within 10 km that participated in any BLM auction during the sample period. The variable ``Manufacturer'' is a dummy variable indicating whether the winner is a sawmill. Winners are frequently manufacturers (sawmills), with a manufacturer share of $\approx 0.62$, indicating asymmetries across buyer types. Although the exact number of sawmills and loggers near the auctioned tracts is unknown, it is likely that loggers outnumber sawmills in this industry.\footnote{In Oregon, approximately 700 logging companies exist (\hyperlink{https://www.dnb.com/business-directory/company-information.logging.us.oregon.html}{Dun \& Bradstreet}, accessed December 13, 2024), compared to around 100 sawmills (\hyperlink{https://www.whereorg.com/businesses/sawmills-or}{Whereorg.com}, accessed December 13, 2024)} Reserve prices average about \$0.59 million and winning bids about \$0.75 million, with the second–highest to highest bid ratio around $B_2/B_1 \approx 0.98$, consistent with meaningful competition.\footnote{This high ratio suggests either strong correlation in bidder values (consistent with common-value elements) or intense competition driving bids close to valuations, both of which are consistent with the timber auction setting where tract characteristics are publicly observable and prices are correlated across bidders of the same type. We nevertheless follow the timber-auction literature (e.g., \citealp{athey2011comparing,haile2001auctions}) and maintain an independent private-values environment to remain comparable with prior empirical work while keeping the dynamic participation and capacity choices in our model tractable.} The variable ``Lumpsum'' is a dummy variable that takes value one if the payment does not depend on actual volumes of cut timber but estimated timber volumes. Most sales are lump‐sum contracts (share $\approx 0.86$). Tracts sold at oral auctions are economically sizable, with mean estimated volume around $4{,}000$ mbf ($\approx 6{,}800$ ccf) and about $230$ acres. These moments point to active competition over sizeable tracts, with systematic heterogeneity across buyer types.

\begin{figure}[h]
\begin{center}
\includegraphics[height = 0.4\textheight]{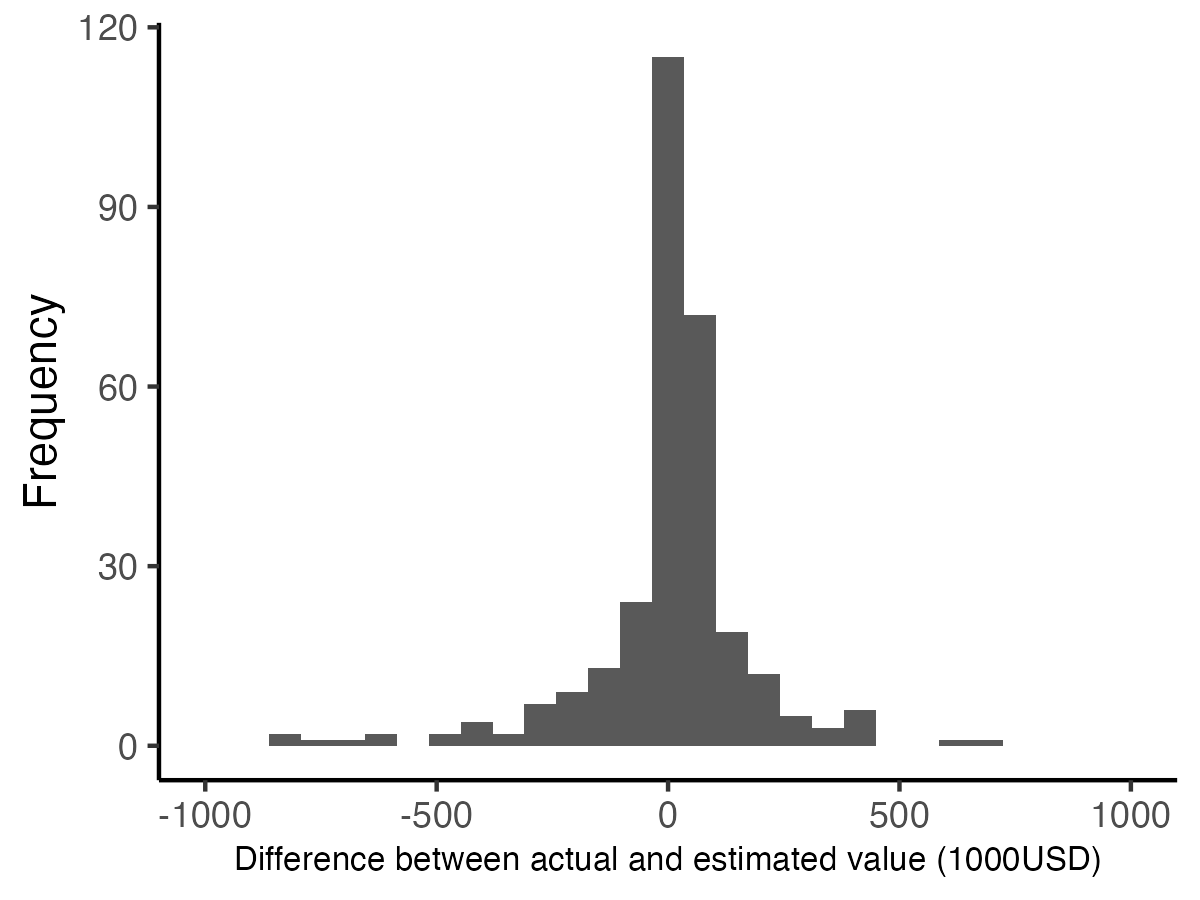}
\end{center}
\caption{The difference between actual and estimated cut timber values}\footnotesize
\label{fg:diff_actualestimated}
\textit{Note}: We use the oral auction data from 2012Q2 to 2023Q1 by the Bureau of Land Management (BLM), U.S. Department of the Interior.
\end{figure}

\paragraph{Prediction accuracy of timber quantities}
Table \ref{tb:summary_statistics_of_diff_actualestimated} shows that actual cutting outcomes closely track ex ante estimates in oral auctions: mean (actual $-$ estimated)/estimated differences are near zero across measures—acres $-0.11$, ccf $-0.03$, mbf $-0.02$, and value $-0.03$. Consistent with these moments, Figure \ref{fg:diff_actualestimated} displays a distribution of value differences tightly centered at zero with modest tails. This close alignment supports the common use of lump‐sum contracts, indicating that auction‐stage estimates provide reliable payment benchmarks for most sales.

\paragraph{Cutting stage}
Panel (b) of Table \ref{tb:summary_statistics_of_entry_and_bidding_stage_oral} summarizes post‐award outcomes in the BLM records for oral auctions. Cutting occurs frequently over a lease, averaging $14.59$ activities. The observed removal timing (“Cutting delay’’) averages $21.99$ months, while the average contract term is $32.43$ months, suggesting that timber removals concentrate toward the end of the lease. We return to the timing of removals in Section \ref{subsubsec:fact_findings}. The “Revised contract term’’ averages $36.72$ months, implying an extension of roughly four months relative to the original term. However, it remains unclear whether this extension significantly benefits buyers in terms of their operational or financial outcomes. Note that the maximum number of “Cutting delay’’ exceeds that of “Revised contract term’’, implying that the raw data includes some glitch. We can avoid this issue by using only 36-month auctions in our estimation.

\begin{figure}[h]
\begin{center}
\includegraphics[height = 0.4\textheight]{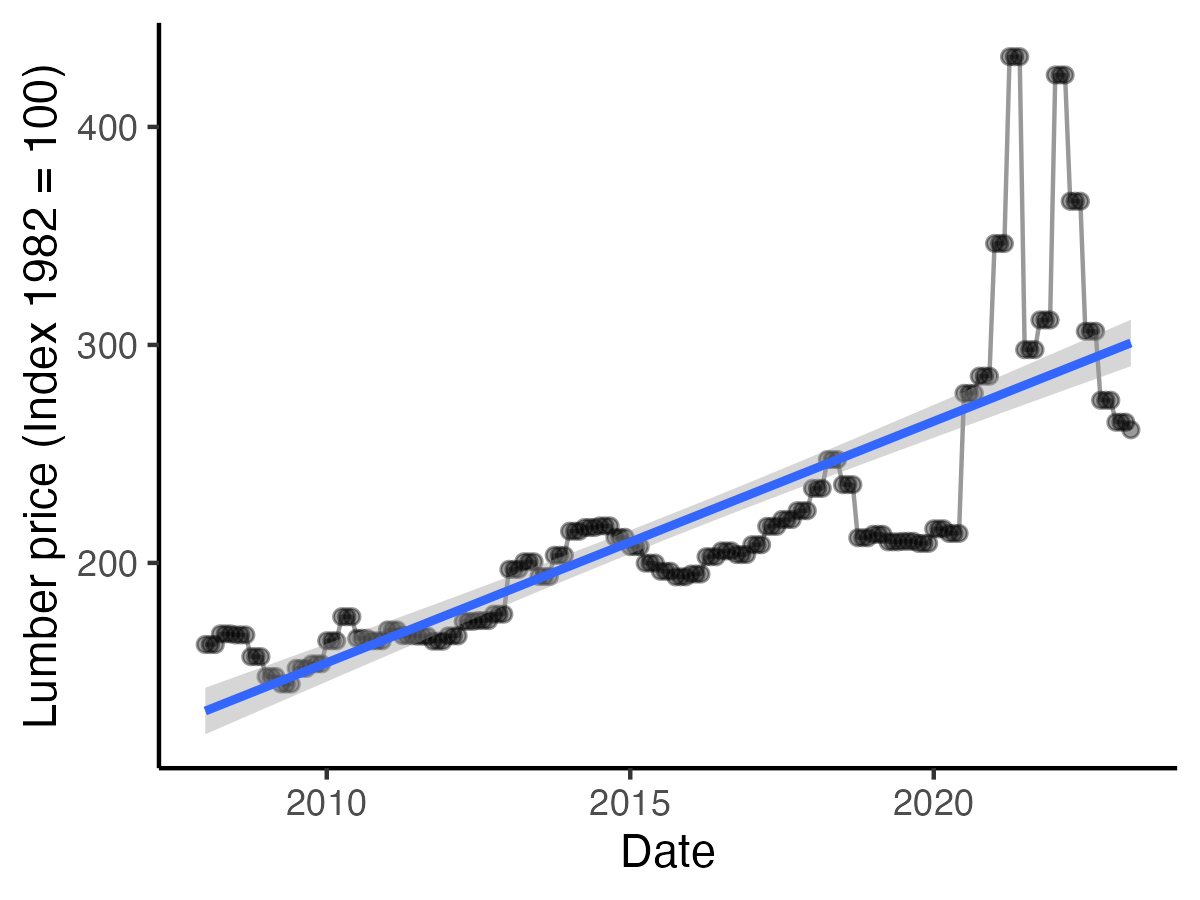}
\end{center}
\caption{Lumber price trends}\footnotesize
\textit{Sources}: Federal Reserve Economic Data (FRED).
\label{fg:lumber_price_trend}
\end{figure}

\paragraph{Lumber price trends}
Figure $\ref{fg:lumber_price_trend}$ describes the time-trend of lumber market prices. The number in this figure represents the index value for lumber market prices. One can observe that lumber prices fluctuate across periods, especially during the COVID-19 pandemic (after 2020). During this period, housing prices soared because people had to stay home and had strong incentives to invest in their houses.  

\begin{table}[ht!]\small
\begin{centering}
      \caption{Summary statistics}      \label{tb:summary_statistics_of_entry_and_bidding_stage_oral} 
      \subfloat[Entry and bidding stage]{\input{figuretable/summary_statistics_of_entry_and_bidding_stage_oral}}\\ 
      \subfloat[Cutting stage]{\input{figuretable/summary_statistics_of_cutting_stage_oral}}\\
\end{centering}
 \footnotesize
  \textit{Note}: We use the oral auction data from 2012Q2 to 2023Q1 by the Bureau of Land Management (BLM), U.S. Department of the Interior. Winning bids ($B_1$) are averaged over all 306 oral auctions, whereas second-highest bids ($B_2$) and the bid-ratio $B_2/B_1$ are reported only for auctions with at least two bidders.
\end{table} 

\begin{table}[h]\small
  \begin{center}
      \caption{Summary statistics of difference between estimate and actual cuts}
      \label{tb:summary_statistics_of_diff_actualestimated} \input{figuretable/summary_statistics_of_diff_actualestimated}
  \end{center}\footnotesize
  \textit{Note}: We use the oral auction data from 2012Q2 to 2023Q1 by the Bureau of Land Management (BLM), U.S. Department of the Interior.
  This table displays the difference between actual and estimated cutting amounts, standardized by the predicted quantities at the auction stage. The table reports basic statistics for four measures: acres, ccf (hundred cubic feet), mbf (thousand board feet), and transaction value. Column labels indicate the physical units of the estimated quantity, but each entry equals (Actual $-$ Estimated)/Estimated in that unit. 
\end{table}

\subsection{Fact findings}\label{subsubsec:fact_findings}
We highlight three empirical patterns that motivate the policy counterfactuals and structural model. First, longer leases appear to raise bidder valuations, as documented with model-free evidence and regressions. Second, completion timing clusters near the lease expiration, pointing to the need for a dynamic model of harvesting. Third, bidder entry shows little sensitivity to auction characteristics, allowing us to treat participation as exogenous in the structural framework.

\paragraph{Effects of lease duration on bidders' value}
Most contracts in our sample feature a 36-month duration (Table $\ref{tb:summary_statistics_of_entry_and_bidding_stage_oral}$), but this benchmark has been debated. In response to public comments that emphasized buyers’ need for more time, BLM began piloting 48-month contracts in 2023.\footnote{See Section 5463.1 of the \href{https://www.federalregister.gov/documents/2020/12/18/2020-27580/forest-management-decision-protest-process-and-timber-sale-administration}{Forest Management Decision Protest Process and Timber Sale Administration} (accessed July 7, 2024).}

\begin{figure}[h]
\begin{center}
\includegraphics[height = 0.4\textheight]{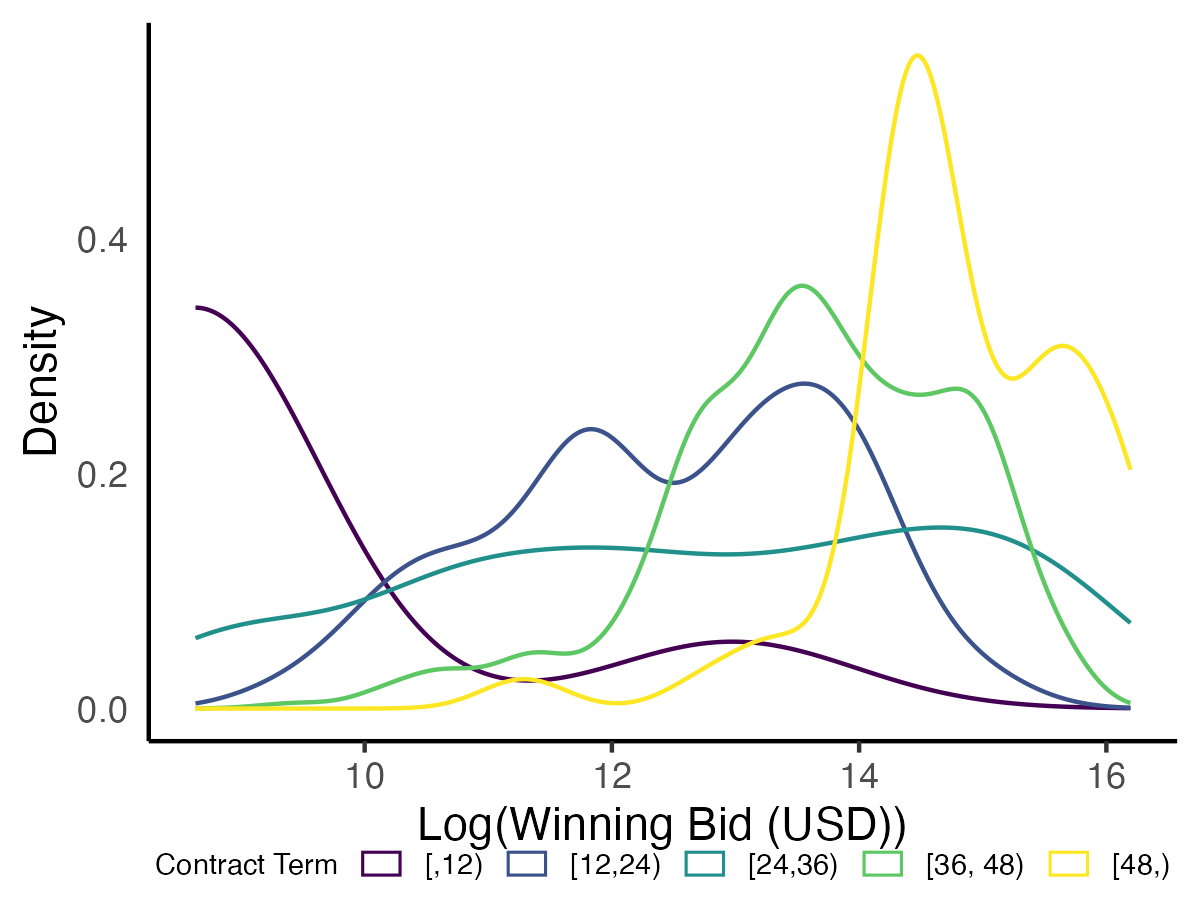}
\end{center}
\caption{Distribution of Winning Bids Conditional On Contract Term}\footnotesize
\textit{Sources}: We use the oral auction data from 2023Q1 to 2025Q2. 
\label{fg:density_log_winning_price_vs_contract_term_2023_2025}
\end{figure}

Because the policy is new, the sample of 48-month contracts is still small. Even so, the model-free evidence in Figure $\ref{fg:density_log_winning_price_vs_contract_term_2023_2025}$ shows a rightward shift in the distribution of winning bids for longer terms. We therefore turn to regression evidence that controls for tract characteristics to test whether duration affects valuations.

Table \ref{tb:regression_result_log_highest_bid_reserve_price} reports regressions of log winning bids and reserve prices on observable auction characteristics. Winning bids for oral auctions rise sharply with contract length, whereas sealed-bid auctions exhibit a weak negative relationship. For oral auctions, adding one month to the term raises the winning bid by roughly 5\%, while the corresponding effect for sealed-bid auctions is a reduction of at most 0.9\%. Hence, contract duration explains an important share of outcome heterogeneity beyond lumber prices, the number of active bidders, reserve prices, and tract size (acres or mbf), echoing the findings in \cite{lu2008estimating}. We revisit the mechanism through counterfactual simulations.

\paragraph{Timing of removal completion}

\begin{figure}[h!]
  \begin{center}
 \subfloat[$25\%$ of progress]{\includegraphics[keepaspectratio, scale=0.8]{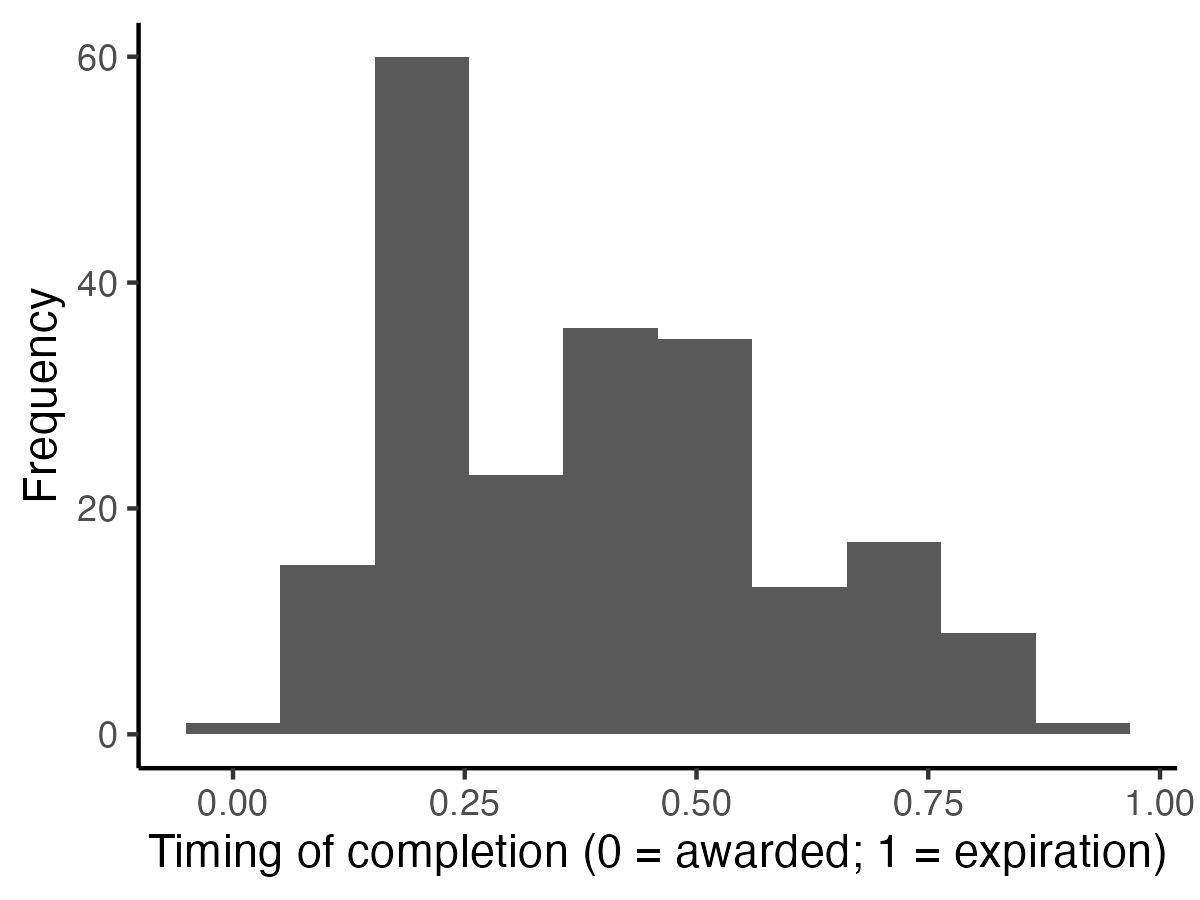}}
  \subfloat[$50\%$ of progress]{\includegraphics[keepaspectratio, scale=0.8]{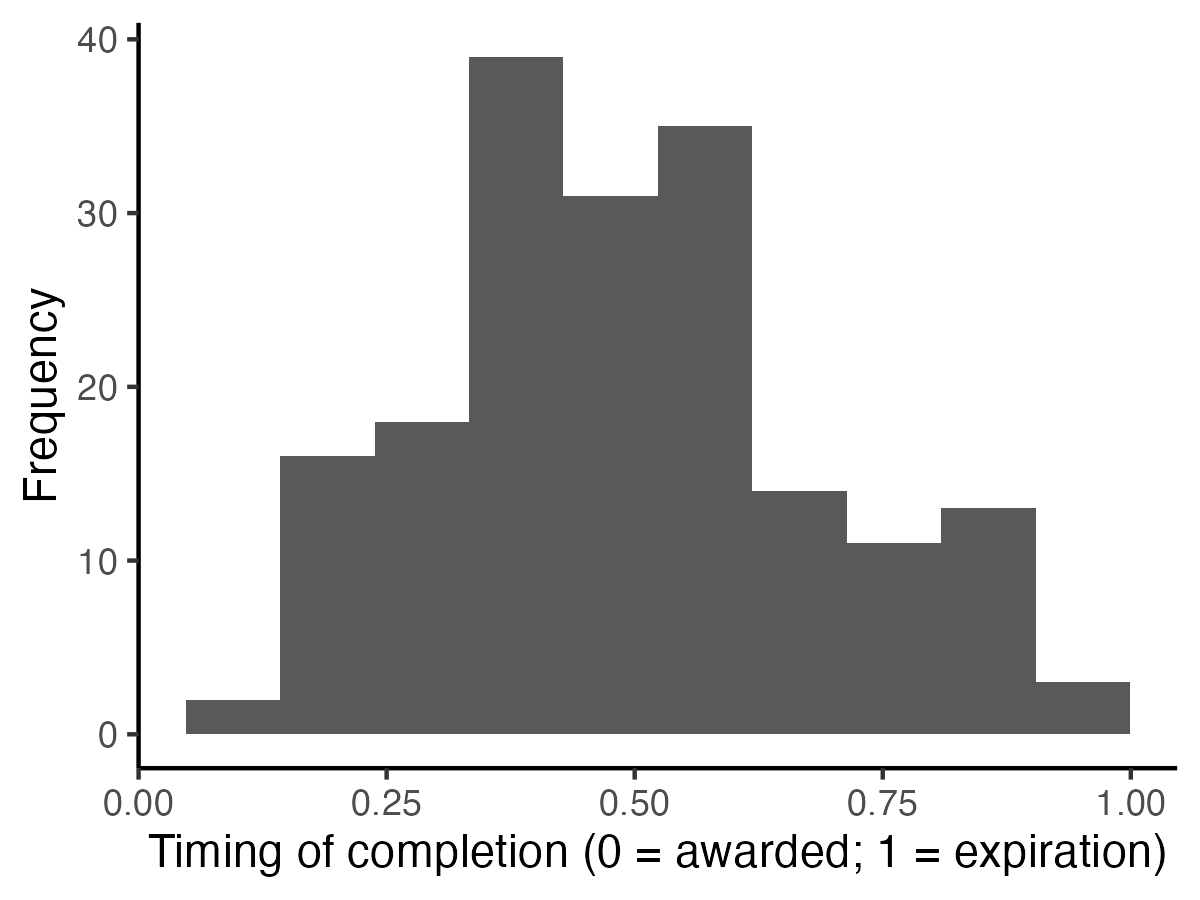}}\\
  \subfloat[$75\%$ of progress]{\includegraphics[keepaspectratio, scale=0.8]{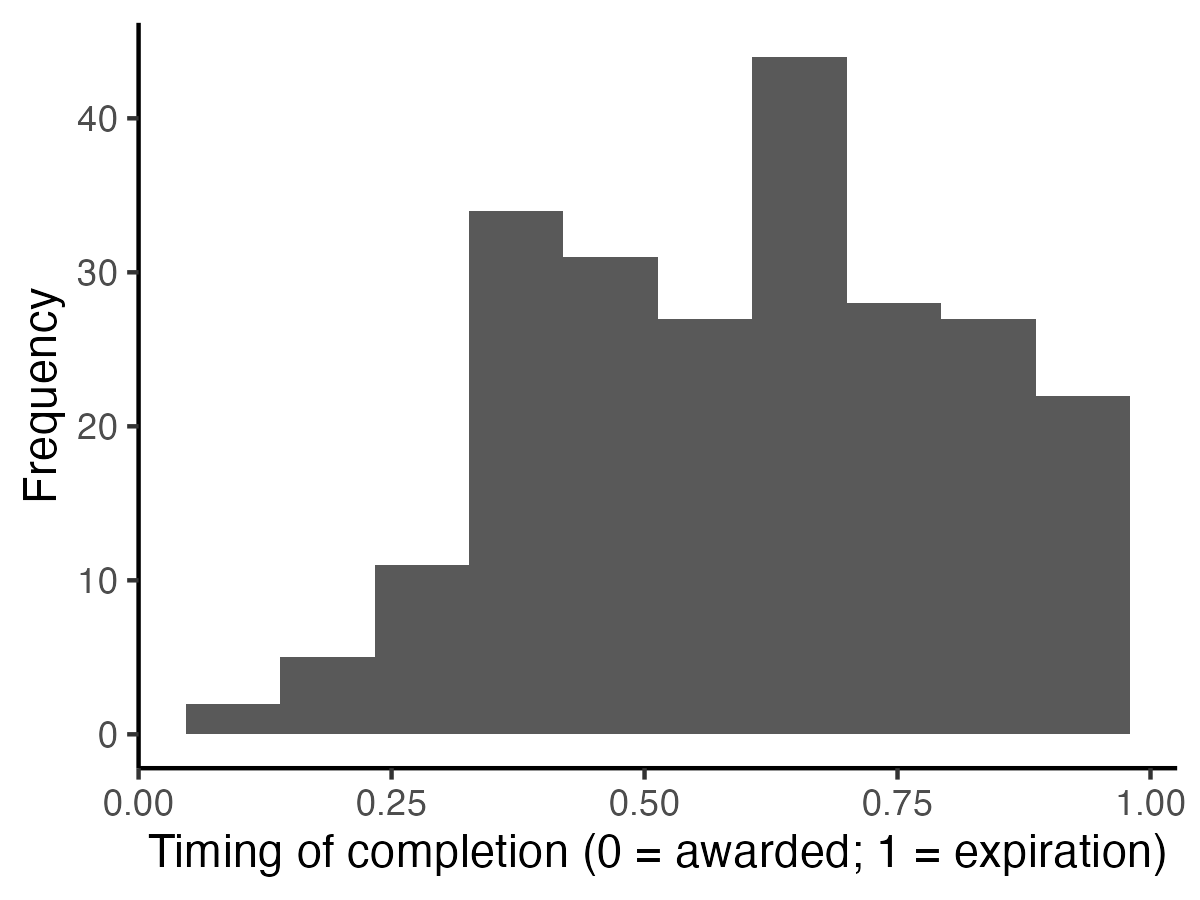}}
  \subfloat[$100\%$ of progress]{\includegraphics[keepaspectratio, scale=0.8]{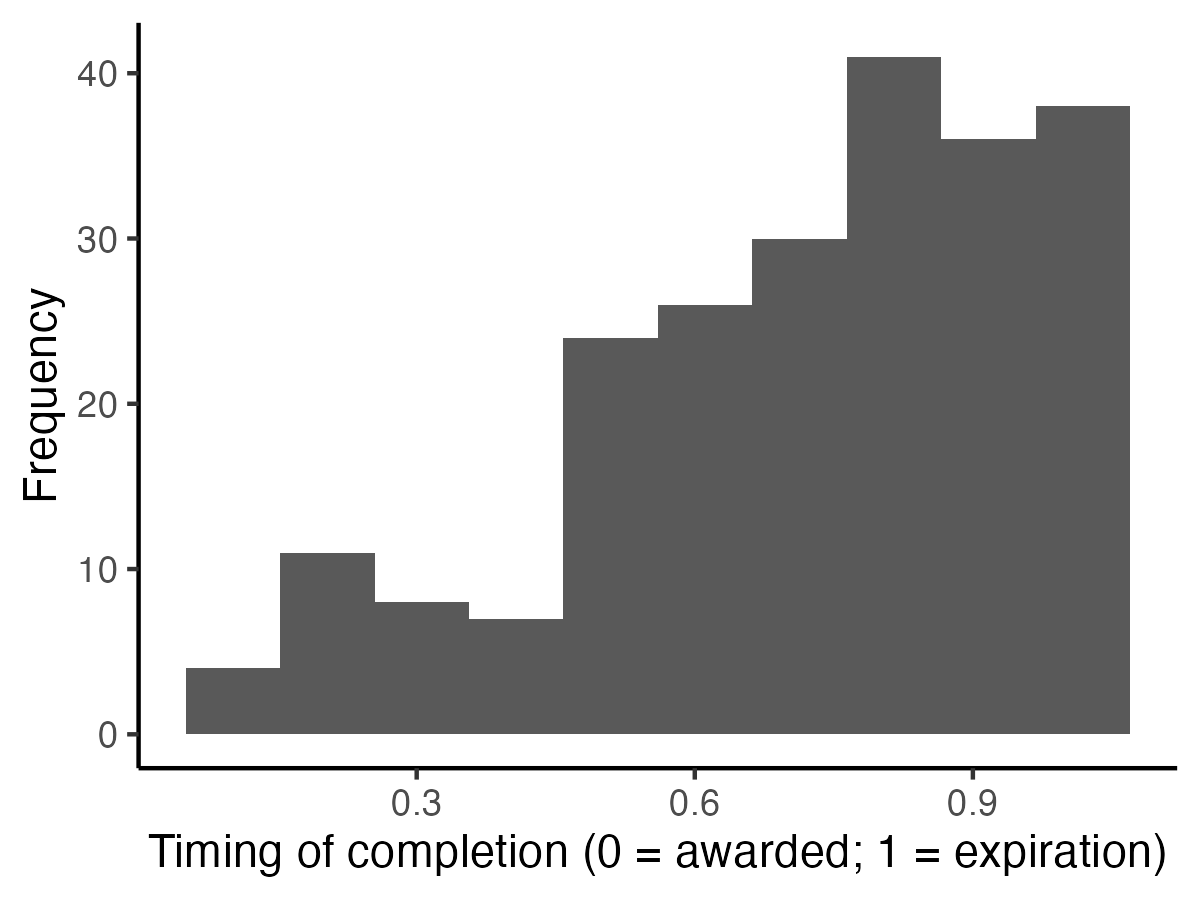}}
  \caption{Distribution of cutting progress}
  \label{fg:time_hist_100} 
  \end{center}
  \footnotesize
   Note: These figures illustrate histograms of the timing of timber removal completion as a proportion of the contract period, where 0 indicates the start of the contract (awarded date) and 1 represents the end of the contract (expiration date). Panels (a)–(d) show the distribution of completion timing for 25\%, 50\%, 75\%, and 100\% progress in timber removal, respectively. 
\end{figure} 

Figure \ref{fg:time_hist_100} summarizes when buyers reach each cutting milestone. Panel (a) shows that even halfway through the contract many tracts have only 25\% harvested. As progress thresholds rise, the modal completion time moves toward the contract end (Panels (b) and (c)). Panel (d) indicates that final completion is heavily concentrated near expiration, underscoring how bidders back-load harvesting to manage price risk and operational constraints.

To study what drives harvesting, we regress quarter-level cutting amounts on quarter-level lumber prices and their interactions with buyer type. We consider three dependent variables—mbf, acres, and value—where mbf denotes one thousand board feet. Table \ref{tb:regression_result_transaction_cut_amount} shows a strong positive relationship between lumber prices and each cutting measure, indicating that buyers accelerate harvesting when prices spike. The coefficients also reveal systematic differences across types: conditional on price and tract size, sawmills harvest larger volumes per quarter than loggers.

\begin{table}[h!]
  \begin{center}
      \caption{Regressions of quarter-level cutting amounts}
      \label{tb:regression_result_transaction_cut_amount} 
      \input{figuretable/regression_result_transaction_cut_amount}
  \end{center}\footnotesize
  Note: This table presents regression results for quarter-level cutting amounts. The dependent variables are measured in three units: ``Transaction mbf'' represents the volume of timber harvested in thousand board feet (mbf), ``Transaction Acres'' refers to the acreage of timber harvested, and ``Transaction Value'' denotes the monetary value of harvested timber. Explanatory variables include ``Lumber Price'', indicating the market price of timber during the quarter, ``Manufacturer'', a dummy variable equal to 1 if the buyer is a sawmill manufacturer, and controls for the total tract size measured in mbf, acres, and value.
  Standard errors are reported in parentheses. ***, **, and * indicate significance at the 1\%, 5\%, and 10\% levels, respectively. 
\end{table}

\paragraph{Exogenous entry}
We estimate Poisson and linear regression models for the number and share of active bidders using the 204 oral auctions with at least two participants (Table \ref{tb:poisson_result_num_of_active_bidders}). The covariates include lease duration, auction format, lumber prices, a COVID-19 dummy, reserve prices, and tract size. With the exception of reserve prices—which are statistically significant but economically small—none of the variables meaningfully predicts entry. This pattern indicates little systematic self-selection into participation beyond tract attractiveness reflected in reserve prices, consistent with the findings in \cite{lu2008estimating}. We therefore treat entry as exogenous and focus on the post-entry stages.

\begin{table}[h]
  \begin{center}
      \caption{Regressions of the number of active bidders and the percentage of active bidders}
      \label{tb:poisson_result_num_of_active_bidders} 
      \begin{center}
         \subfloat[Poisson]{\input{figuretable/poisson_result_num_of_active_bidders}}
          \subfloat[Linear]
         {\input{figuretable/regression_result_share_of_active_bidder}}
      \end{center}
      
  \end{center}\footnotesize
 Note: This table presents Poisson regression results for the number of active bidders and linear regression results for the percentage of active bidders as dependent variables. ``Contract term'' represents the lease duration in months.
 Additional covariates include ``Lumber price'' for market timber prices, ``Covid dummy'' accounting for the COVID-19 period, ``Reserve price'' indicating the auction's minimum bid, and ``Acres'', representing tract size. Standard errors are reported in parentheses. The analysis is based on 204 oral auctions with at least two bidders, with RMSE values provided for model fit. ***, **, and * indicate significance at the 1\%, 5\%, and 10\% levels, respectively.
\end{table} 

\begin{landscape}
\begin{table}[h]
{\footnotesize
  \begin{center}
      \caption{Regressions of log of winning bids and reserve prices}
      \label{tb:regression_result_log_highest_bid_reserve_price} 
      \input{figuretable/regression_result_log_highest_bid_reserve_price}
  \end{center}}\footnotesize
  Note: This table presents regression results for the natural logarithm of winning bids and reserve prices as dependent variables. Explanatory variables include ``Contract term''
  and other covariates including ``Lumber price'', representing market timber prices; ``Covid dummy'', a variable accounting for the COVID-19 period; ``Num of active bidders'', the number of participants actively bidding; ``Acres'' and ``Mbf'', which capture tract size in acres and thousand board feet, respectively; and ``Manufacturer'', a dummy variable equal to 1 if the buyer is a sawmill manufacturer. The analysis is based on 204 oral auctions with at least two bidders. Standard errors are reported in parentheses. ***, **, and * indicate significance at the 1\%, 5\%, and 10\% levels, respectively. 
\end{table} 
\end{landscape}

\section{Model}\label{sec:model}
In this section, we propose a structural model and discuss its equilibrium outcome. Our model has two sequential stages: the bidding stage and the cutting stage. In the bidding stage, both loggers and sawmill manufacturers submit their bids. Then, the winning buyer in the bidding stage obtains the right to consume timber, and she faces a single-agent dynamic problem: when and how much timber to cut until the lease expires. In the following Section \ref{subsec:setup}, we provide more details of our model settings and introduce notations. In Section \ref{subsec:analysis}, we derive solutions to our model, which we bring to data in the following estimation part. 

\subsection{Setup}\label{subsec:setup}

\paragraph{Bidding stage}
We model each oral auction $k$ as having $n_k$ bidders, taking participation as exogenous. The auction characteristics $\boldsymbol{x}_k$ include the contract term $T_k$, the initial timber volume $u_{k0}$, and the current lumber price $p_0$. Bidders belong to two observable types—loggers ($l$) and sawmills ($s$). Bidder $i$’s valuation is
$v_{ik} = \xi_{ik} V^{m(i)}_0(\boldsymbol{x}_k)$, the product of an idiosyncratic component $\xi_{ik}$ and a common component $V^{m(i)}_0(\boldsymbol{x}_k)$ determined by the cutting-stage problem for type $m(i)\in\{l,s\}$. The term $\xi_{ik}$ captures private information such as operational efficiency or opportunity cost and follows type-specific distributions $F_l(\cdot)$ or $F_s(\cdot)$. Because timber quantities and prices are common to all bidders, values are common within each type up to $\xi_{ik}$. In an oral ascending auction, bidders successively raise the standing price until only one bidder remains. This final bidder wins the item and pays the final price.

\paragraph{Cutting stage}
After winning auction $k$, the buyer faces a finite-horizon single-agent problem that spans $T_k$ periods. At the beginning of each period $t \in \{1,2,\ldots,T_k\}$, winner $i$ observes the state $s_{kt} = (p_t, u_{kt})$, which consists of the lumber price and the remaining volume of timber. The initial state $s_{k0}$ coincides with the auction characteristics $\boldsymbol{x}_{k}$, so $u_{k0}$ equals the estimated timber volume at the time of sale. Given the state, the winner chooses how much timber to harvest. The decision variable $q_{kt} \in \{0, 0.25, 0.5, 0.75, 1\}$ denotes the fraction of the \textit{original} tract cut in period $t$, so the physical quantity harvested in that period is $q_{kt} u_{k0}$. This discretization keeps the dynamic choice problem tractable while capturing the salient variation in cutting behavior, striking a balance between matching observed removal patterns and avoiding the curse of dimensionality that a finer grid would trigger. For example, $q_{kt}=0.5$ means the winner cuts half of the original tract in that period. The state transitions satisfy
\begin{align}
    u_{k,t+1} &= u_{kt} - u_{k0}q_{kt};  \\
    p_{t+1} &\sim f(\cdot|p_{t}) ,
\end{align}
where $f(\cdot|p_{t})$ is the conditional distribution of next period’s lumber price. Because $u_{kt}$ records the remaining physical inventory, it satisfies $u_{kt}=u_{k0}\bigl(1-\sum_{\tau < t} q_{k\tau}\bigr)$. We impose $u_{k1}=u_{k0}$ because the estimated timber volume aligns closely with realized volume (Section \ref{sec:data}). We also treat lumber prices as exogenous to individual auctions, as BLM sales comprise a small share of total supply.

Each period, the winner earns the following payoff from cutting a share $q_{kt}$ of the tract:
\begin{align}
    q_{kt}u_{k0}(\gamma_{m(i)}p_t - c_{1m(i)} - c_{2m(i)}q_{kt}u_{k0})+\varepsilon^{q_{kt}}_{ikt},
\end{align}
where $\gamma_{m(i)}$ captures type-specific price sensitivity, $c_{1m(i)}$ is the marginal cutting cost, and $c_{2m(i)}$ is the curvature term. The quadratic term allows for increasing marginal costs when concentrating harvesting within a period—capturing equipment-capacity and labor constraints—while remaining flexible enough to encompass the linear case. The last term $\varepsilon^{q_{kt}}_{ikt}$ is an action-specific shock. Because the tract must be fully harvested, the choices satisfy $\sum_{t=1}^{T_{k}} q_{kt} = 1$.

\subsection{Analysis}\label{subsec:analysis}
We solve the model backward. First, we characterize the winner’s dynamic program and derive the continuation value at the initial period. Second, we plug this continuation value into the bidding problem.


\paragraph{Cutting stage}
In the cutting stage, winner $i$ in auction $k$ aims to maximize her expected future payoff. In other words, she solves the following Bellman equation:

\begin{align}\label{eq:bellman_equation}
    V^{m(i)}_t(s_{kt}, \varepsilon_{ikt}) =& \max_{q_{kt}}q_{kt}u_{k0}(\gamma_{m(i)}p_t - c_{1m(i)} - c_{2m(i)}q_{kt}u_{k0}) + \varepsilon^{q_{kt}}_{ikt} \\
    &+  \beta E_{\varepsilon_{ikt+1}}\left[V^{m(i)}_{t+1}(s_{k,t+1}, \varepsilon_{ikt+1}|s_{kt}, \varepsilon_{ikt},q_{kt})\right], \nonumber\\
    & \text{subject to }\begin{cases}
    q_{kt} \leq 1 - \sum^{t-1}_{\tau=1}q_{k\tau},& \text{ if }t<T_k \\
    q_{kt} = 1 - \sum^{t-1}_{\tau=1}q_{k\tau},& \text{ if }t=T_k 
    \end{cases},\nonumber
\end{align} 
where $\beta$ is the discount factor. Because all timber must be removed by expiration, the terminal value satisfies $V^{m(i)}_T(s_{kT}, \varepsilon_{ikT}) = u_{kT}(\gamma_{m(i)}p_T - c_{1m(i)} - c_{2m(i)}u_{kT})$. The first two terms in Equation \eqref{eq:bellman_equation} deliver the period payoff from cutting $q_{kt}$, while the third term captures the expected continuation value, integrating over next period’s lumber price. The constraints enforce full harvest within the lease. Following \cite{rust1987optimal}, we impose conditional independence, $\Pr(s_{kt+1}, \varepsilon_{ikt+1}\mid s_{kt}, \varepsilon_{ikt}, q_{kt}) = \Pr(s_{kt+1}\mid s_{kt}, q_{kt}) \Pr(\varepsilon_{ikt+1}\mid s_{kt})$, and assume the shocks $\varepsilon^{q_{kt}}_{ikt}$ are drawn from type-I extreme value distribution for each action $q_{kt}$.
Then, using the part of Equation \eqref{eq:bellman_equation}, we define the integrated value function as
\begin{align}
    V^{m(i)}_t(s_{kt}) =& \int\biggl[\max_{q_{kt}}q_{kt}u_{k0}(\gamma_{m(i)}p_t - c_{1m(i)} - c_{2m(i)}q_{kt}u_{k0}) + \varepsilon^{q_{kt}}_{ikt} \nonumber\\
    &+  \beta E_{\varepsilon_{ikt+1}}\left[V^{m(i)}_{t+1}(s_{k,t+1}|s_{kt},q_{kt})\right]\biggr] f(\varepsilon_{ikt}) d\varepsilon_{ikt}, \nonumber\\
    =&\int[v_{ikt}^{q_{kt}}(s_{kt}) + \varepsilon^{q_{kt}}_{ikt}] f(\varepsilon_{ikt}) d\varepsilon_{ikt},\nonumber
\end{align}
where $v_{ikt}^{q_{kt}}(s_{kt})$ is the choice-specific continuation value.
Finally, we obtain the conditional choice probability that the winner chooses $q_{kt}$ as follows:
\begin{align}
    \operatorname{Pr}(q_{kt}|s_{kt}) &=  \frac{\exp\left(v_{ikt}^{q_{kt}}(s_{kt})\right)}{\sum_{q'_{kt}}\exp\left(v_{ikt}^{q'_{kt}}(s_{kt}))\right)}.
\end{align}
The continuation value at the initial period $V^{m(i)}_0(s_{k0})$ determines the common part of bidders' value discussed in the previous section. 

\paragraph{Bidding stage}
Given the continuation value at the initial period $V^{m(i)}_0(\boldsymbol{x}_k)$, the bidding strategy at equilibrium is straightforward, as we consider oral auctions. All bidders have a dominant strategy of bidding up to their valuation: $b^*_{ik}=\xi_{ik}V^{m(i)}_0(\boldsymbol{x}_k)$. Therefore, the payment by the winner at equilibrium is equivalent to the second-highest value among the participants. 

\section{Estimation}\label{sec:estimation}
This section identifies and estimates the structural parameters. We distinguish between dynamic parameters—price sensitivities and cutting costs—and auction parameters governing private values and type shares. Recovering both sets is essential for the counterfactuals. In our estimation, we focus on the auctions with a three-year contract duration because different contract durations require us to derive the different finite-horizon continuation value, which leads to computational burden. 

The BLM auction data report the number of bidders, the winner’s identity and type, and the winning bid for each oral auction. Using these records and information on nearby firms, we infer the pool of potential loggers and sawmills. The cutting records provide quarterly harvest amounts for each tract, and FRED supplies quarterly lumber prices. Variation in prices pins down the price sensitivities of loggers and sawmills, while observed harvest quantities identify marginal costs. Once we estimate the dynamic primitives, we can compute tract-level continuation values. Combining those with winning bids identifies the distribution of private values. Finally, variation in the counts of potential and actual bidders reveals the composition of bidder types, allowing us to avoid relying on bid variation alone as in \cite{kong2022identification}.

The baseline model assumes type-specific parameters are homogeneous within each type. Yet in practice even sawmills differ markedly in inventory capacity and downstream demand, implying heterogeneity in cutting costs and valuation responses. We therefore extend the framework to recover unobserved heterogeneity using the random-coefficients method of \cite{fox2011simple-bef}.

\subsection{Estimation strategies}\label{sec:estimation_strategy_benchmark}
\subsubsection{Dynamic parameters}
\paragraph{Cutting stage}
The cutting decision is a discrete choice over reduction levels $\{0,0.25,0.5,0.75,1\}$, which match the quarter-level aggregates observed in the data. Using the dynamic discrete-choice model, we estimate each type’s price sensitivity $\gamma_m$ and cost parameters $\boldsymbol{c}_m$ with the Nested Fixed Point Algorithm of \cite{rust1987optimal}. The inner loop computes continuation values and choice probabilities backward from the terminal period for a given parameter guess, and the outer loop updates the parameters to maximize the likelihood of observed cutting paths.

\subsubsection{Auction parameters}
\paragraph{Entry stage}
Let $\lambda_m$ denote the entry probability for type $m\in\{l,s\}$, collected in $\boldsymbol{\lambda}=(\lambda_l,\lambda_s)$. These probabilities pin down the composition of entrants because the share of loggers equals $\lambda_l/(\lambda_l+\lambda_s)$. Using the observed number of participants $n_k$ in auction $k$ and external information on the pool of potential bidders $N_k^m$,\footnote{Potential bidder counts come from negotiation records and Oregon Department of Forestry data.} we evaluate
\begin{align}
    \Pr(n_k|\boldsymbol{\lambda}, n_k>0) = \sum_{j =0}^{n_k}\frac{\Pr(j \text{ loggers}|N^l_k,\lambda_l)\Pr(n_k-j \text{ sawmills}|N^s_k,\lambda_s)}{1-\Pr(\text{no logger})\Pr(\text{no sawmill})}.\label{eq:likelihood_entry}
\end{align}
The numerator gives the probability that $j$ loggers and $n_k-j$ sawmills enter, while the denominator conditions on the auction not being canceled for lack of bidders (which we do not observe). Each probability is computed from a Poisson distribution with mean $\lambda_m$, and we estimate $\boldsymbol{\lambda}$ by maximizing the resulting log-likelihood.

\paragraph{Bidding stage}
Given the dynamic parameters, we estimate the valuation distributions using the observed winning bids. Let bidder $i$ of type $m\in\{l,s\}$ win an $N$-bidder auction at price $\tau$. Following \cite{kong2022identification}, the likelihood of this event is
{\footnotesize
\begin{align}
    p(i \text{ wins with }\tau | N,V^{m}_{0}; \hat{p}) &= \left(1-F^{m(i)}(\tau|V^{m(i)}_{0}) \right)\sum_{n=0}^{N-1} {N-1 \choose n} \hat{p}^n (1-\hat{p})^{N-1-n} \nonumber\\
    &\quad \times \sum_{j =1}^{N-1} \Bigl(1[j \leq n] f^l_j(\tau|V^{l}_{0})[F^l(\tau|V^{l}_{0})]^{n-1}[F^s(\tau|V^{s}_{0})]^{N-1-n} \nonumber\\
    &\qquad\qquad + 1[j > n] f^s_j(\tau|V^{s}_{0})[F^l(\tau|V^{l}_{0})]^{n}[F^s(\tau|V^{s}_{0})]^{N-2-n}\Bigr),
\end{align}
}
where $\hat{p}:=\hat{\lambda}_l/\hat{\lambda}_s$ is the entrants share estimated above and $V^{m(i)}_{0}$ is the continuation value for $i$'s type implied by the dynamic model. The last bracket in the equation represents the case where either a logger or a sawmill submits the second-highest bid. For each type of $m$, we specify $F^{m}(\tau|V^{m}_{0})$ as a gamma distribution with scale and shape $(\mu_{m}V^{m}_{0},\sigma_{m})$, a flexible choice that accommodates the skewness seen in the bid data and fits the empirical bid distribution well. We estimate $(\mu_{m},\sigma_{m})$.

\subsection{Estimation strategy for random coefficients}\label{subsec:estimation_appendix}

To estimate heterogeneous primitives, we employ the method of \cite{fox2011simple-bef}. They interpret random-coefficient models as mixtures over a finite grid of coefficient “types”, which makes outcome probabilities linear in type weights. Note that these mixture types need not coincide with the observed logger–sawmill classification; we pool bidders before the algorithm assigns weights. We compute each type’s model-implied probabilities once, and then estimate the mixture weights by maximizing the likelihood subject to positivity and the simplex constraint (weights are nonnegative and sum to one). This approach avoids imposing a parametric distribution on heterogeneity while remaining computationally tractable.

To implement the approach, we index the grid of parameter vectors by $r$, where each vector $\theta^{r} = (\gamma^{r}, c_{1}^{r}, c_{2}^{r})$ captures a potential coefficient type in the pooled sample. Let $\omega^{r}$ denote the weight on type $r$; estimating the random-coefficient model reduces to estimating these weights. For comparability with the baseline estimates, we hold $(c_{1}^{r}, c_{2}^{r})$ fixed at the sawmill parameters reported in Table \ref{tb:estimation_results} and allow heterogeneity to operate through $\gamma^{r}$.

We solve for the weights $\boldsymbol{\omega}=(\omega^1,\ldots,\omega^R)$ by minimizing
\begin{align}
    \min_{\{\omega^r\}} \sum_{i}^{N_{m(i)}}\sum_{j \in \{0,.25,.5,.75,1\}}\sum_{t=1}^{13}\left(q_{ijt} - \boldsymbol{z}^\top_{ijt}\boldsymbol{\omega}\right)^2
\end{align}
subject to $\omega^r \ge 0$ for all $r$, $\sum_r \omega^r = 1$, and
$F_{\tilde{v}}(\cdot|s) =  F_{V_0}(\cdot|s)$. Here $q_{ijt}$ is an indicator for choosing action $j$ at time $t$, and $\boldsymbol{z}_{ijt} = (z^1_{ijt},\ldots,z^R_{ijt})$ stacks the conditional choice probabilities $\Pr(q_{it}=j|s_{it}; \theta^r)$ implied by each grid point. Unlike \cite{fox2011simple-bef}, our finite-horizon setting requires indexing these probabilities by $t$.

The last constraint equates the distribution of the highest values $\tilde{v}$ to that implied by the continuation values $V_0(s;\theta^r)$. Because we observe only the winning bid—the second-highest value in an oral auction—we recover $F_{\tilde{v}}$ using order statistics before estimation.

We assume bidder valuations follow $f(\cdot|s; \mu)$, where $\mu$ parameterizes the distribution. For each auction $k$ in state $s$ we observe the number of bidders $n_k$ and the winning price $p_k$. Because $p_k$ equals the second-highest value in an oral auction, its density is

\begin{align}
    \underbrace{\text{Pr}(p_k=p)}_{\text{data}} = \underbrace{n_k(n_k-1)f(p|s; \mu)(1-F(p|s; \mu))F(p|s; \mu)^{n_k-2}}_{\text{order-statistics model}}.
\end{align}
We estimate $\mu$ via maximum likelihood:
\begin{align}
    \max_\mu \sum_k \log\text{Pr}(p_k=p).
\end{align}
Recovering $f(\cdot|s; \mu)$ thus delivers the distribution of highest values $f_{\tilde{v}}(\cdot|s; \mu)$, which we match to the distribution of continuation values at $t=0$.

\paragraph{Identification}
Identification in \cite{fox2011simple-bef} requires the $IJT \times R$ matrix $\boldsymbol{Z}=\{z_{ijt}\}$ to have rank $R$. We therefore keep the number of grid points modest: enlarging $R$ can make the conditional choice probabilities nearly collinear, especially because our data exhibit bunching in removal timing. To satisfy the rank condition we focus on heterogeneity in $\gamma$, holding $c_1$ and $c_2$ at the estimates in Section \ref{sec:constant_param_for_each_bidder_type}.

\subsection{Monte Carlo simulation}
Because our estimation procedure extends the standard approach, we evaluate its performance with 500 Monte Carlo replications.

For the auction component, we simulate 500 auctions under the model in Section \ref{sec:model}, fixing $V^{m}_{0}=1$ so that $\mu_m$ captures the scale parameter $\mu_m V^{m}_{0}$. The true parameters are $(\mu_{l}, \sigma_l, \mu_s, \sigma_s)=(1,1,2,3)$ and $(\lambda_l, \lambda_s) = (0.1,0.15)$, governing the gamma value distributions and entry probabilities, respectively.

For the dynamic component, we simulate 1,000 agents solving the cutting problem with $p_t \in \{1,\ldots,10\}$ and transitions $p_{t+1}\sim f(p|\mu=p_t,\sigma^2=1)$. The discount factor is $\beta=0.95$, and the true parameters are $(\gamma, c_1, c_2)=(1,0.5,0.05)$.

Panel (a) of Table \ref{tb:estimated_auction_bias_rmse} reports the bias and RMSE for the auction parameters, showing good performance aside from a mild upward bias in the sawmill scale parameter. Panel (b) reports the dynamic-parameter diagnostics, which reveal small downward bias in the price-sensitivity estimate.

\begin{table}[!htbp]
  \begin{center}
      \caption{Monte Carlo simulation results}
      \label{tb:estimated_auction_bias_rmse} 
      \subfloat[Auction parameters]{\input{figuretable/auction_param_bias_rmse}}
      \subfloat[Dynamic parameters]{\input{figuretable/montecarlo_dynamic_bias_rmse}}
      
  \end{center}\footnotesize
  \textit{Sources}: 500 simulations. For auction parameters, we set $(\mu_{l}, \sigma_l, \mu_s, \sigma_s)=(1,1,2,3)$, and $(\lambda_l = 0.1, \lambda_s = 0.15)$ which represent shape and scale parameters for gamma distribution for each type of bidders, and $\lambda$ captures entry rate. For dynamic parameters, we set $(\gamma, c_1, c_2)=(1,0.5,0.05)$, which represent price sensitivity and cutting cost per each MBF and its square value, respectively.
\end{table}

\section{Estimation Results}\label{sec:estimation_results}

We begin by estimating the dynamic parameters separately for each group of bidders, classified by their observed types—loggers and sawmills. Next, using the estimated parameters as a benchmark, we estimate the random-coefficient model based on \cite{fox2011simple-bef}, allowing for random coefficients regarding the timber prices, which would be the most critical factor in the bidder's dynamic decision.

\subsection{Constant parameters for each bidder type}\label{sec:constant_param_for_each_bidder_type}

\begin{table}[!htbp]
  \begin{center}
      \caption{Estimation results}
      \label{tb:estimation_results} 
      \subfloat[Cutting stage]{\input{figuretable/cutting_stage_estimate}}
      \subfloat[Bidding stage]{\input{figuretable/bidding_stage_estimate}}
      \subfloat[Entry stage]{\input{figuretable/entry_stage_estimate_oral}}
  \end{center}\footnotesize
  \textit{Note}: Estimates of parameters underlying each stage. $(\gamma_m,c_{1m},c_{2m})$ represent price sensitivity and removal costs in the dynamic stage for type $m$. $(\mu_m, \sigma_m)$ are the scale and shape parameters of the idiosyncratic value distribution, and $\lambda_m$ denotes the entry probability for type $m$. Standard errors are based on 1,000 bootstrap samples drawn from the auction data.
\end{table} 

\paragraph{Cutting stage}

Panel (a) in Table \ref{tb:estimation_results} presents the estimation results of dynamic parameters at the cutting stage. The findings suggest that sawmills exhibit lower sensitivity to lumber prices compared to loggers, as evidenced by the parameter \(\gamma_s = 0.038\) being smaller than \(\gamma_l = 0.140\). This indicates that loggers are more inclined to increase timber cutting in response to rising prices, whereas sawmills adjust their cutting activity less frequently in response to such price changes. The difference is likely due to the fact that loggers, often small businesses, lack sufficient storage capacity for harvested timber, making them more responsive to high lumber prices as an incentive to cut timber.

Additionally, the estimates indicate that loggers tend to incur higher cutting costs (\(c_{1l}=25.405\)) compared to manufacturers (\(c_{1s}=7.242\)). This disparity can likely be attributed to differences in business size and structure. Manufacturers, who integrate logging, delivery, and processing operations, can achieve efficiencies across these stages, thereby reducing overall cutting costs. In contrast, loggers—typically smaller, independent businesses—must transport cut timber to their customers' processing facilities, introducing an additional layer of logistical inefficiency. This lack of integration adds to their operational costs, exacerbating the cost disparity between loggers and larger manufacturers. The curvature terms are also consistent with this pattern but not statistically significant, with \(c_{2l}=0.005\) and \(c_{2s}\approx 0.000\).

Overall, these findings underline the importance of accounting for bidder heterogeneity when modeling timber auctions. Differences in price sensitivity and operational costs play a critical role in shaping bidders’ continuation values, which directly influence bidding strategies in the auction stage. These dynamics highlight the distinct dynamic considerations faced by loggers and sawmills in both cutting and auction contexts.

\paragraph{Bidding stage}

Panel (b) of Table \ref{tb:estimation_results} presents the parameters of the gamma distributions governing the idiosyncratic factors $\xi$. Sawmills exhibit both higher means and greater dispersion than loggers: $(\mu_s,\sigma_s)=(8.166,1.065)$ versus $(\mu_l,\sigma_l)=(0.988,0.768)$. Because the variance of a gamma distribution equals $\mu^2 \sigma$, sawmills display substantially more heterogeneity, consistent with evidence from \cite{athey2011comparing} that they are stronger and more varied bidders. The large $\mu_s$ also indicates that dynamic considerations alone cannot explain sawmills' WTP, whereas the logger mean $E[\xi_l]=0.988\times0.768\approx 0.759$ is relatively close to one compared to the sawmill mean $E[\xi_s]=8.166\times1.065\approx 8.697$, so logger bids align more closely with the continuation value from the cutting problem.

\begin{figure}[!ht]
  \begin{center}
  \subfloat[Bidding stage]{\includegraphics[keepaspectratio, scale=0.7]{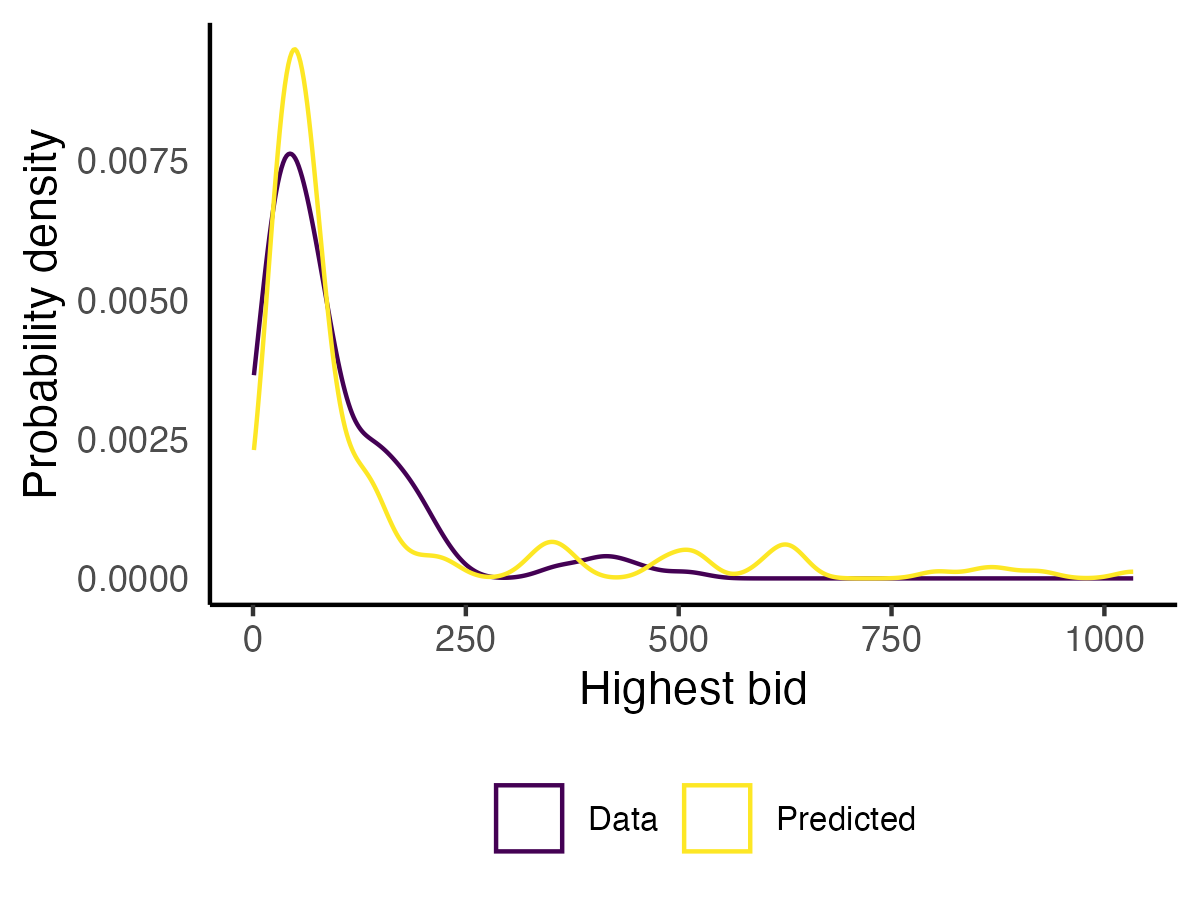}}
  \subfloat[Entry stage]{\includegraphics[keepaspectratio, scale=0.7]{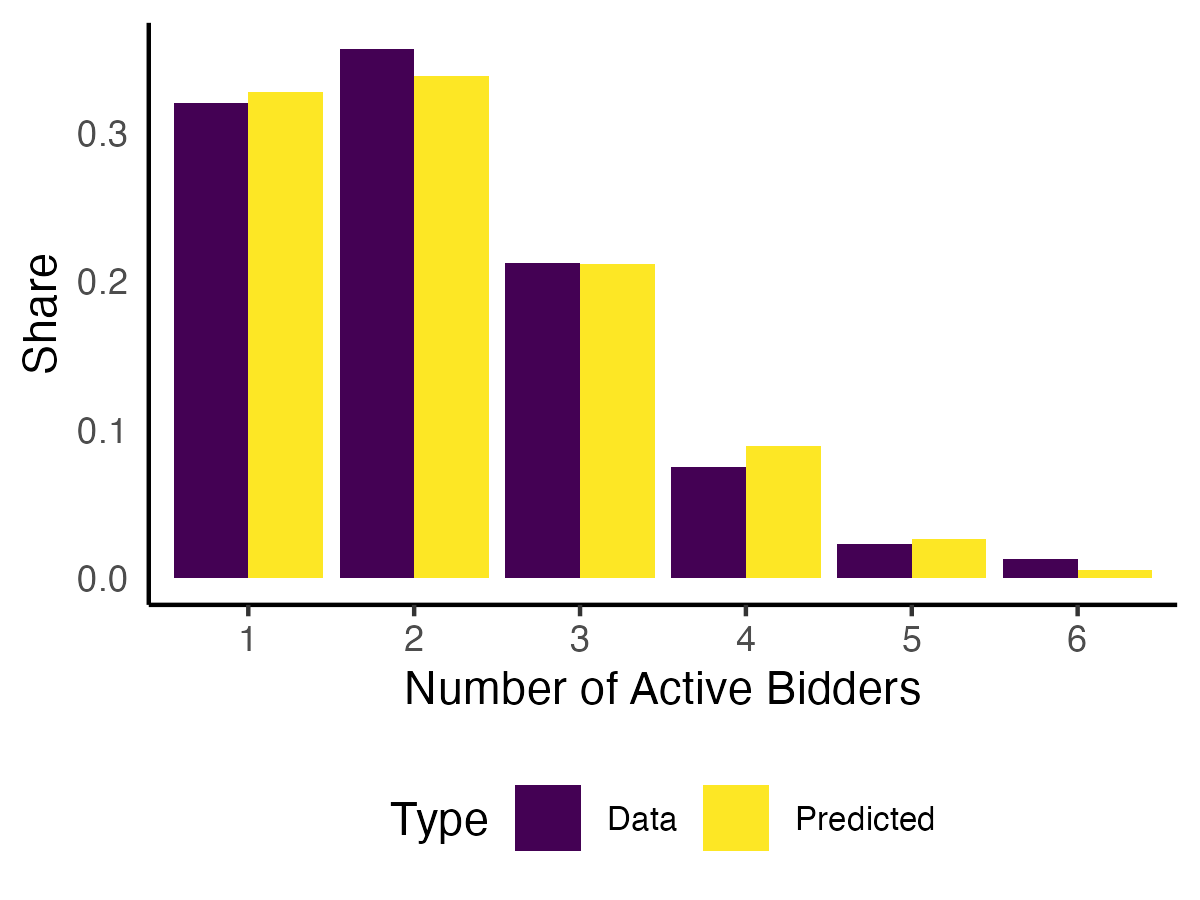}}
  \caption{Model validation}
  \label{fg:model_fit_bidder_valuation_highest_bid} 
  \end{center}
  \footnotesize
  Note: These figures validate the model by comparing predicted outcomes with observed data. Panel (a) presents the probability density of the highest bids in the bidding stage, where the purple line represents the actual data, and the yellow line indicates the model-predicted distribution. Panel (b) shows the distribution of the number of active bidders in the entry stage, with observed data in purple and model predictions in yellow. 
\end{figure} 

Panel (a) in Figure \ref{fg:model_fit_bidder_valuation_highest_bid} illustrates the probability density of the highest bids in the data and prediction using estimated parameters. Both histograms display a similar pattern, with winning prices clustering around $120$ and exhibiting a rightward skew.

\paragraph{Entry stage}

Panel (c) of Table \ref{tb:estimation_results} reports similar entry probabilities for loggers and sawmills, implying comparable propensities to participate. Combined with the higher win rates for sawmills, this pattern reinforces the view that sawmills are stronger bidders rather than simply more likely to show up. Panel (b) of Figure \ref{fg:model_fit_bidder_valuation_highest_bid} shows that the estimated model replicates the observed distribution of active bidders.

\subsection{Heterogeneous parameters for both bidder types}

\begin{figure}[h!]
  \begin{center}
  \includegraphics[keepaspectratio, scale=1.2]{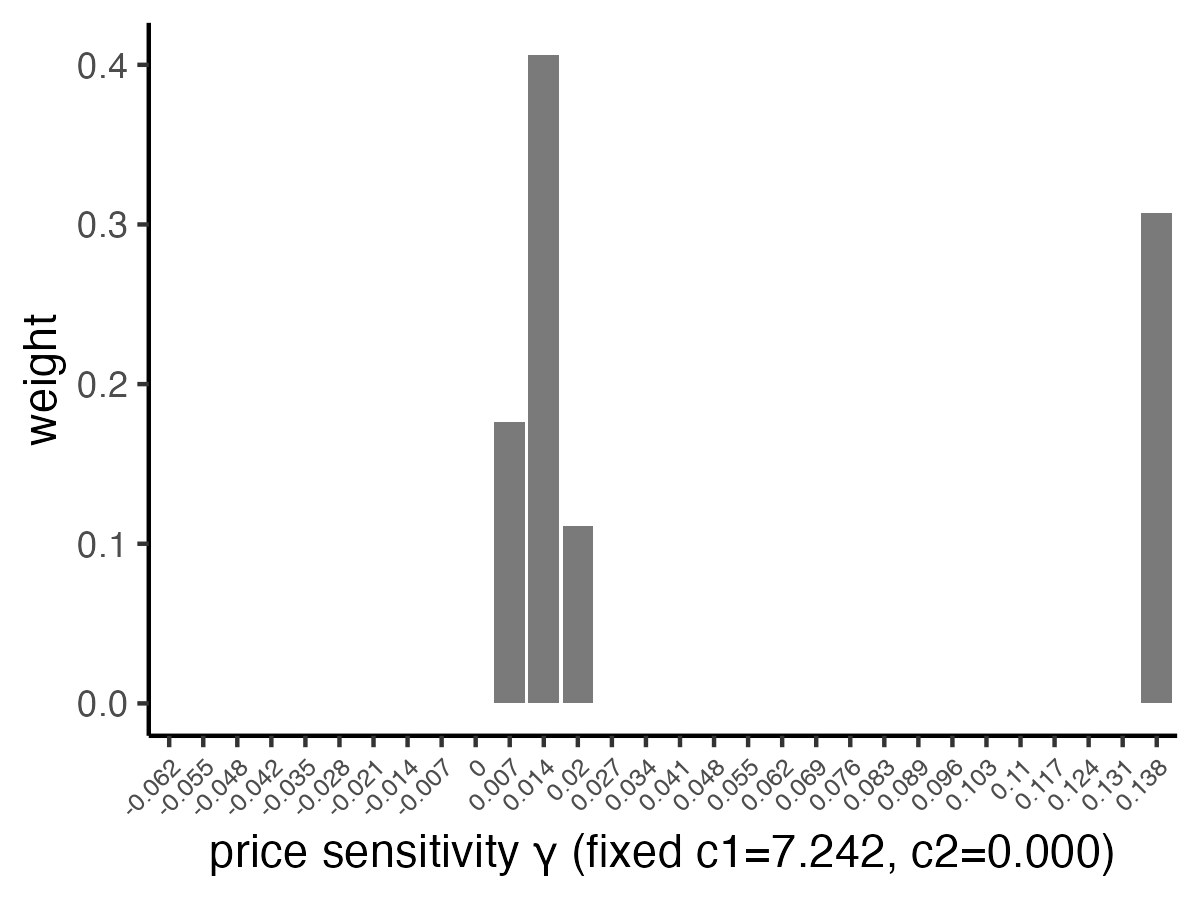}
  \caption{Estimated distributions of heterogeneity for the price sensitivity and cutting cost}
  \label{fg:parameter_estimate_fkbr} 
  \end{center}
  \footnotesize
   Note: We take 30 grids on the domain of the price sensitivity parameter $\gamma$ and fix $c_{1}$ and $c_{2}$ to the sawmill parameters reported in Panel (a) of Table \ref{tb:estimation_results}.
\end{figure}

In Section \ref{sec:constant_param_for_each_bidder_type}, we assumed that type-specific dynamic parameters are homogeneous within each type, which may be restrictive. To relax this, Figure \ref{fg:parameter_estimate_fkbr} reports estimates from a random-coefficient specification that allows bidder-level heterogeneity in price sensitivity without pre-classifying bidders as loggers or sawmills. The resulting distribution exhibits two clear masses close to the parameter values reported in Panel (a) of Table \ref{tb:estimation_results}, indicating that while meaningful heterogeneity is present, most of the variation in price sensitivity is effectively captured by the logger–sawmill split. In short, the richer model confirms the baseline pattern: allowing for random coefficients refines the fit but does not overturn the conclusion that the primary axis of variation aligns with bidder type.

\section{Counterfactual}\label{sec:counterfactual}

\subsection{Effects of contract length on cutting value}\label{subsec:counterfactual_continuation_value}

\begin{figure}[!ht]
  \begin{center}
 \subfloat[Logger's continuation value conditional on price grid level = 1, 2]{\includegraphics[keepaspectratio, scale=0.8]{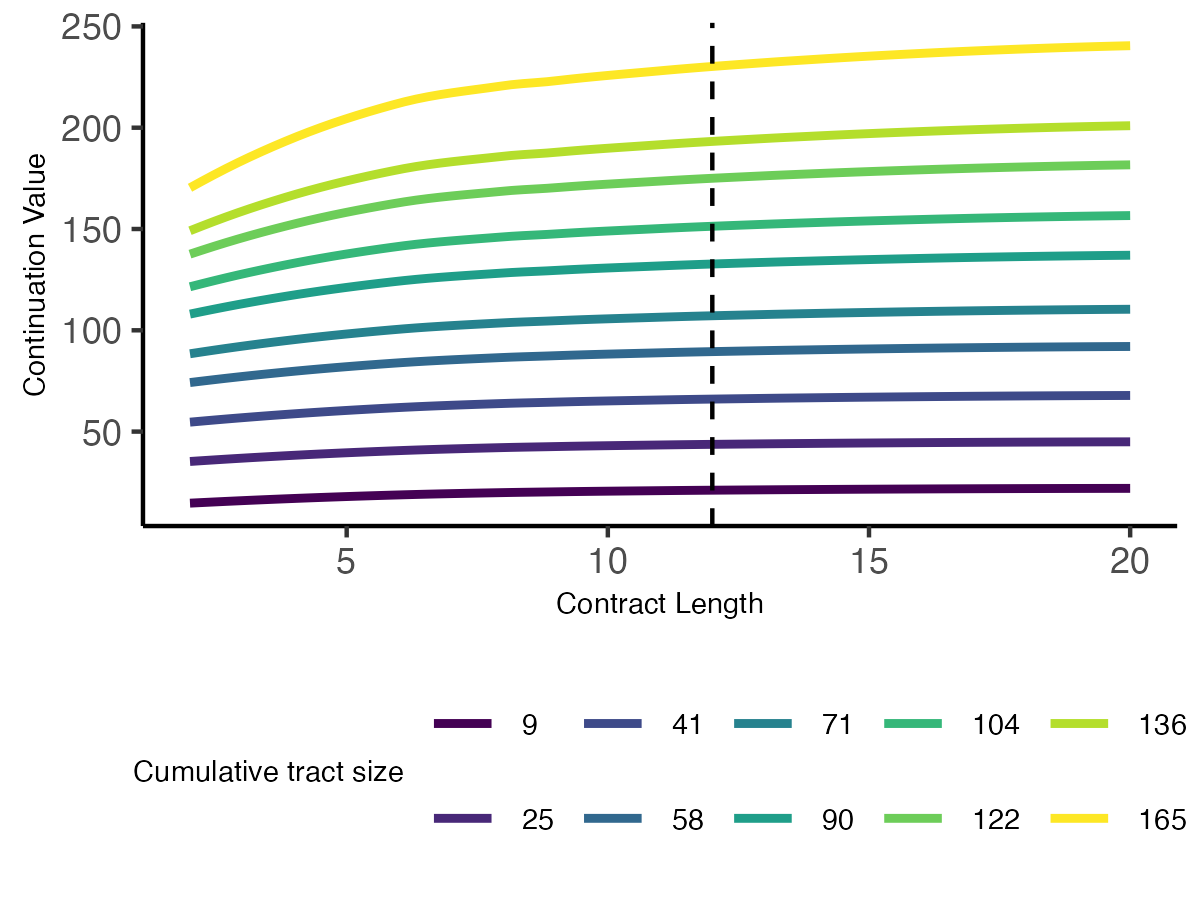}\includegraphics[keepaspectratio, scale=0.8]{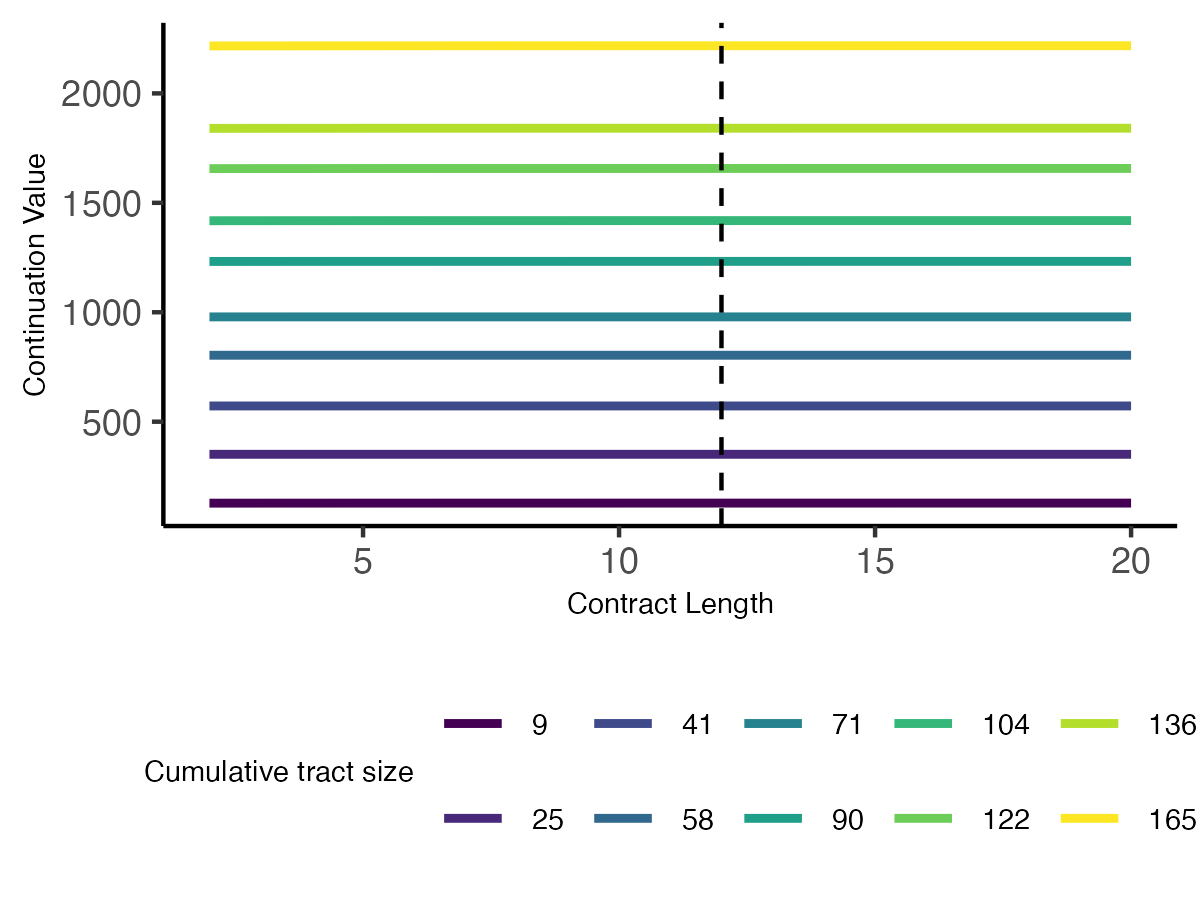}}\\
  \subfloat[Sawmill's continuation value conditional on price grid level = 1, 2]{\includegraphics[keepaspectratio, scale=0.8]{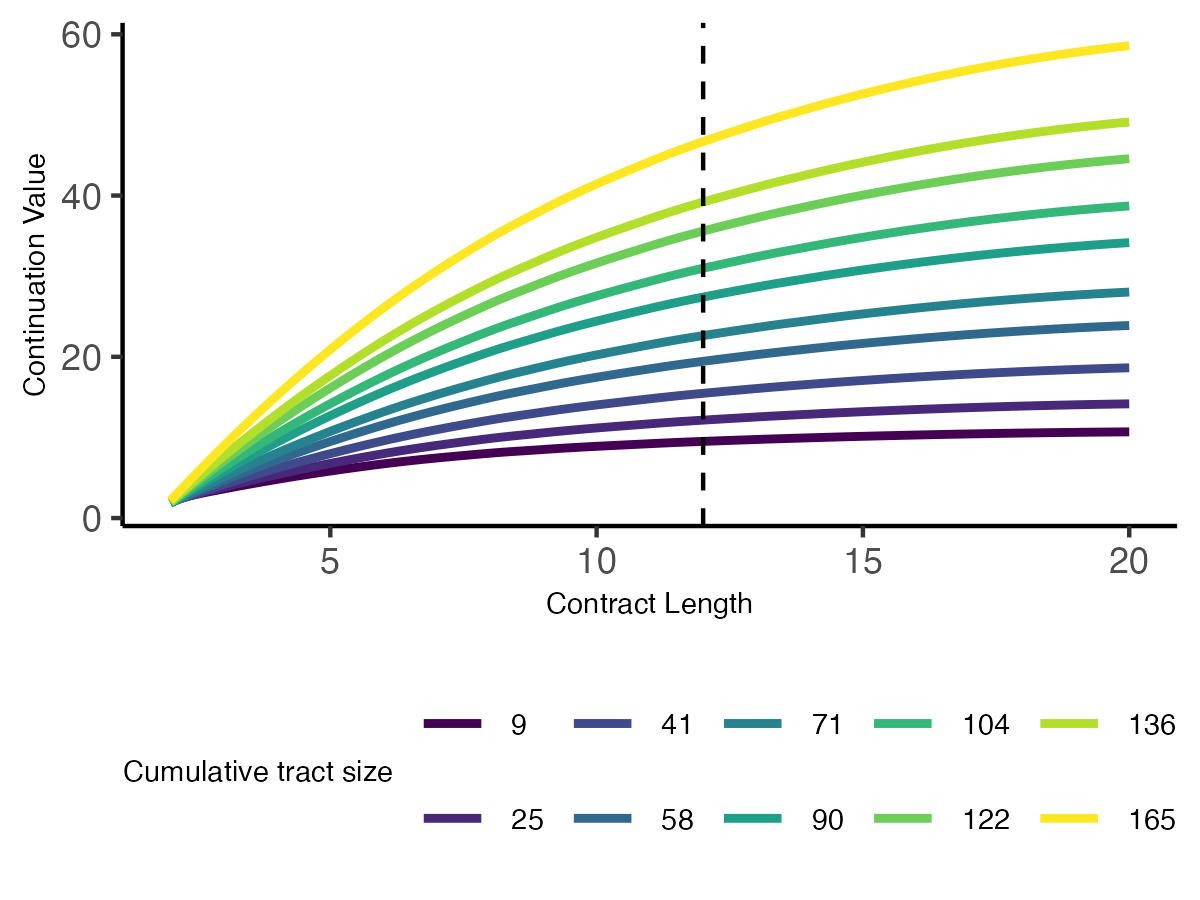}\includegraphics[keepaspectratio, scale=0.8]{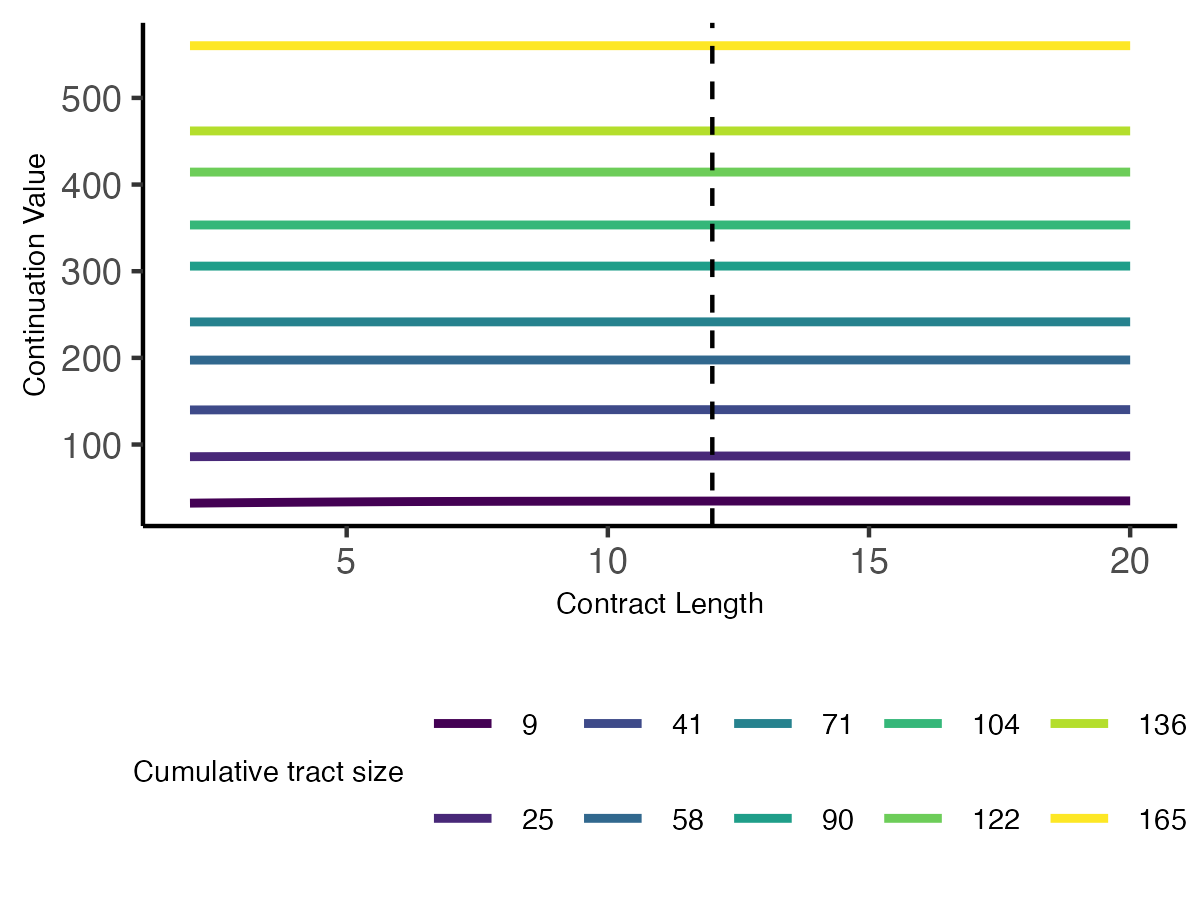}}
  \caption{Continuation value for each contract length}
  \label{fg:continuation_value_plot} 
  \end{center}
  \footnotesize
  Note: This figure shows how the estimated continuation values in the units of $\$1000$ vary by contract length, holding either a low (price grid = 1) or intermediate (price grid = 2) lumber price environment at the time of bidding. Each colored line represents a different tract size. Panel (a) focuses on loggers, while panel (b) presents the corresponding values for sawmills. The vertical dashed line indicates the baseline contract length observed in our data. For exposition we plot the panels using their own vertical scales, so the apparent slope differences reflect both levels and curvature.
\end{figure} 

Before proceeding to the full auction simulations, we begin by examining the baseline continuation values that underpin bidders’ valuations at the time of bidding. Figure \ref{fg:continuation_value_plot}, panel (a), displays the continuation values for loggers across varying contract lengths and tract sizes, under two distinct lumber price states (price grid level 1 for lower prices and level 2 for intermediate prices at the time of bidding). Figure \ref{fg:continuation_value_plot}, panel (b), presents the corresponding continuation values for sawmills under similar conditions. In both panels, each line represents a different cumulative tract size, illustrating how resource scale interacts with contract length and price conditions.

Several key patterns emerge from these results. First, as shown in panel (a) for loggers, extending the contract length consistently enhances continuation values, particularly in low initial price scenarios (price grid = 1). When initial prices are subdued, the option to wait for more favorable market conditions is highly valuable, causing valuations to rise markedly as the contract horizon lengthens. Under intermediate initial price conditions (price grid = 2), continuation values start from a higher baseline, and while they still increase with contract length, the marginal gains are somewhat more moderate. Panel (b) for sawmills shows a similar overall pattern in levels but with a more concave contract-length profile than for loggers—characterized by relatively steep gains at short horizons and weaker marginal increases as the term becomes longer. This description focuses on the low-price grid (price state = 1); at intermediate prices (grid = 2) the curves shift upward but become almost flat with only a slight concavity. The pattern is consistent with our estimates that mills’ cutting responses to price are less elastic; accordingly, contract extensions primarily elevate the overall level of valuations rather than generating large incremental gains at longer horizons.

Second, tract size amplifies the influence of contract length and price state. Larger tracts command higher continuation values across all horizons, as indicated by the upward shift of curves in both panels. This pattern reflects the increased absolute gain from deferring harvest until prices improve. Greater resource availability broadens strategic options, enabling both loggers and sawmills to better time extraction decisions. Although both bidder types benefit, the shape and steepness of their valuation curves differ, highlighting that their cost structures and market strategies respond differently to temporal and price-based incentives.

Finally, comparing panels (a) and (b) shows that bidder type matters. Sawmills appear more sensitive to contract extensions under low-price scenarios, experiencing sharper increases in continuation values as the horizon lengthens. Loggers also benefit from long-term flexibility, but their valuation growth is generally less steep at the margin and sits at higher levels. These differences suggest that the interplay of contract design, market expectations, and resource endowments varies systematically across bidder types.

In sum, the evidence from Figure \ref{fg:continuation_value_plot} demonstrates that contract length, market state, and tract size jointly shape the baseline valuations that bidders bring into the auction. Longer horizons substantially increase the option value of future market conditions, larger tracts magnify potential payoffs, and distinct bidder types respond in unique ways. These findings set the stage for our subsequent counterfactual simulations, where we investigate how these underlying valuation patterns influence equilibrium bidding behavior and ultimately affect auction outcomes.

\subsection{Effects of contract length on transaction price}
Next, we combine the simulated continuation values from Section \ref{subsec:counterfactual_continuation_value} with the estimated auction primitives to quantify how contract duration affects transaction prices. We consider scenarios that vary tract size, bidder composition (and counts), and contract length.

\begin{table}[!htbp]
  \begin{center}
      \caption{Counterfactuals: expected revenue in symmetric auctions}
      \label{tb:counterfactual_winning_price_symmetric} 
       \subfloat[Small Tract Size]{\input{figuretable/counterfactual_winning_price_symmetric_small_tract_oral}}\\
      \subfloat[Large Tract Size]{\input{figuretable/counterfactual_winning_price_symmetric_large_tract_oral}}
  \end{center}\footnotesize
  \textit{Note}: Panels (a) and (b) report average revenue for the small- and large-tract counterfactuals, respectively. In the ``Participants'' column, S (L) denotes a sawmill (logger). Columns four through seven list the contract lengths. Values are in thousands of dollars.
\end{table} 

\begin{table}[!htbp]
  \begin{center}
      \caption{Counterfactuals: expected revenue in asymmetric auctions}
      \label{tb:counterfactual_winning_price_asymmetric} 
       \subfloat[Small Tract Size]{\input{figuretable/counterfactual_winning_price_asymmetric_small_tract_oral}}\\
      \subfloat[Large Tract Size]{\input{figuretable/counterfactual_winning_price_asymmetric_large_tract_oral}}
  \end{center}\footnotesize
  \textit{Note}: Panels (a) and (b) report average revenue for the small- and large-tract asymmetric counterfactuals, respectively. In the ``Participants'' column, S (L) denotes a sawmill (logger). Columns four through seven list the contract lengths. Values are in thousands of dollars.
\end{table}

Table \ref{tb:counterfactual_winning_price_symmetric} shows that lengthening the contract duration—for example, from 12 to 16 quarters (three to four years)—raises expected revenue in symmetric environments. With two identical sawmills bidding on a small tract (panel (a)), revenue climbs from roughly 103.17 thousand dollars at 12 quarters to 112.78 at 16 quarters (+9.3\%). Even for the two-logger case, revenue increases from 30.59 to 31.13 (+1.8\%). The magnitudes differ across bidder types, but the mechanism is the same: additional time lets buyers better align harvesting with market conditions, which raises WTP.

Increasing competition among identical bidders dampens the percentage gains from longer contracts. Comparing the (S,S) and (S,S,S) cases in Table \ref{tb:counterfactual_winning_price_symmetric}, panel (a), shows that adding a third sawmill raises revenues at both durations but reduces the relative lift: +9.3\% versus +9.1\%. With more aggressive bidders, much of the value of longer horizons is already priced into shorter contracts, so the marginal benefit shrinks even though the level effect remains positive.

When bidder types differ—as in matches of sawmills (S) and loggers (L)—the symmetric results need not carry over. Loggers gain less from longer horizons, so revenue growth depends on which type sets the second-highest bid. Table \ref{tb:counterfactual_winning_price_asymmetric} reports these asymmetric scenarios.

Panel (a) of Table \ref{tb:counterfactual_winning_price_asymmetric} shows that in an (S,L) match, extending the contract from 12 to 16 quarters lifts revenue by only 1.6\% (33.35 to 33.89). The second-highest value—typically the logger’s—binds in these auctions, so revenue reflects the type with weaker gains from flexibility. By contrast, a (S,S,L) configuration on a small tract delivers a 9.3\% increase (103.17 to 112.78), similar to the fully symmetric sawmill case. Hence, buyer composition, rather than the number of competitors per se, governs how duration translates into revenue.

Tract size is another key boundary condition. Larger tracts magnify the revenue gains from extended contract duration in both symmetric and asymmetric cases. For example, in Table \ref{tb:counterfactual_winning_price_symmetric}, panel (b), the (S,S) scenario on a large tract sees revenue jump from about 413.58 to 470.14 (+13.7\%), compared with +9.3\% on the small tract. The same pattern appears in asymmetric cases (Table \ref{tb:counterfactual_winning_price_asymmetric}, panel (b)), where higher resource value heightens the payoff to timing flexibility.

In summary, these counterfactual results reveal that the positive impact of lengthening the contract duration on expected revenues is robust yet contingent upon multiple factors. The market trend, composition of these buyers and the tract size play key roles in shaping the impacts of extended horizons on final revenue. For managers and policymakers selling the product with real option nature, these insights underscore the importance of aligning contract structures not only with buyer type composition and market thickness but also with the scale of the underlying product and lumber price environment.

\section{Conclusion}\label{sec:conclusion}
This paper quantifies the effect of contract duration on buyer willingness-to-pay (WTP) in B2B contracts that feature a real option. Using data from U.S. timber auctions, we investigate the value of flexibility—specifically, the buyer's right to flexibly time the harvesting of timber over a specified contract period in the face of evolving market uncertainty.

Our empirical analysis confirms that buyers strategically use this flexibility; harvesting decisions are responsive to lumber prices and harvesting activity is significantly concentrated toward the end of the contract term, demonstrating the option-like nature of the contract. The structural estimation highlights significant heterogeneity between buyer types. Sawmills are more sensitive to the market conditions, and are characterized by more dispersed valuations compared to loggers. This heterogeneity is critical for understanding the value of contract length.

We perform counterfactual analyses to evaluate the impact of extending contract duration on seller revenue. Our analysis shows that extending the contract period (e.g., from 12 to 16 quarters) robustly increases the seller's expected revenue by enhancing the real option value for buyers. Furthermore, this effect is contingent on key boundary conditions. The revenue gains from a longer duration are magnified for larger tracts and are most pronounced in growing markets with a sufficient number of high-type potential buyers. These boundary conditions provide actionable guidance for sellers when tailoring contract duration to market composition and project characteristics.

As a managerial implication, our findings suggest that managers should treat contract duration as an active pricing lever rather than a passive administrative term. By extending the time horizon, sellers effectively transfer a valuable risk-management tool to buyers, which can be monetized through higher upfront prices. This mechanism applies broadly to any setting where consumption timing is flexible, such as the expiration terms of customer reward programs that \cite{inman1994coupon}, \cite{shu2010procrastination}, and \cite{sun2019model} have studied. Just as we observe that high-type timber bidders (sawmills) place a higher premium on duration than smaller loggers, managers in other industries must recognize that the value of flexibility is not uniform. For instance, in a loyalty program, high-frequency business travelers may value the option to delay redemption differently than casual leisure travelers. Therefore, optimal contract design requires segmentation: managers should tailor validity periods to match the volatility and risk profiles of specific buyer segments, balancing the revenue lift from increased WTP against the operational liability of keeping contracts open longer.

Our analysis has limitations that also point to natural extensions for future research. While the framework is applicable to other real-option contexts, such as spectrum licenses or carbon allowances, such applications would require estimating context-specific primitives, as provisions and market structures vary. Future modeling extensions could also provide a richer understanding of buyer valuation. For instance, incorporating buyer risk preferences would clarify how much of the option value is tied to managing price uncertainty, rather than purely to profit maximization. Furthermore, our single-asset model abstracts from broader market interactions; a natural next step is to explore how the option's value is interdependent with both a firm's internal portfolio and the external strategic behavior of rival firms. Finally, we model contracts as fixed agreements, but future work could investigate endogenous renegotiation to understand how the possibility of ex-post bargaining impacts ex-ante valuation.

\bibliographystyle{aer}
\bibliography{timber_auction_bib}

\newpage
\appendix

\section{Data Appendix}\label{sec:data_appendix}

\begin{figure}[h]
\begin{center}
\includegraphics[height = 0.4\textheight]{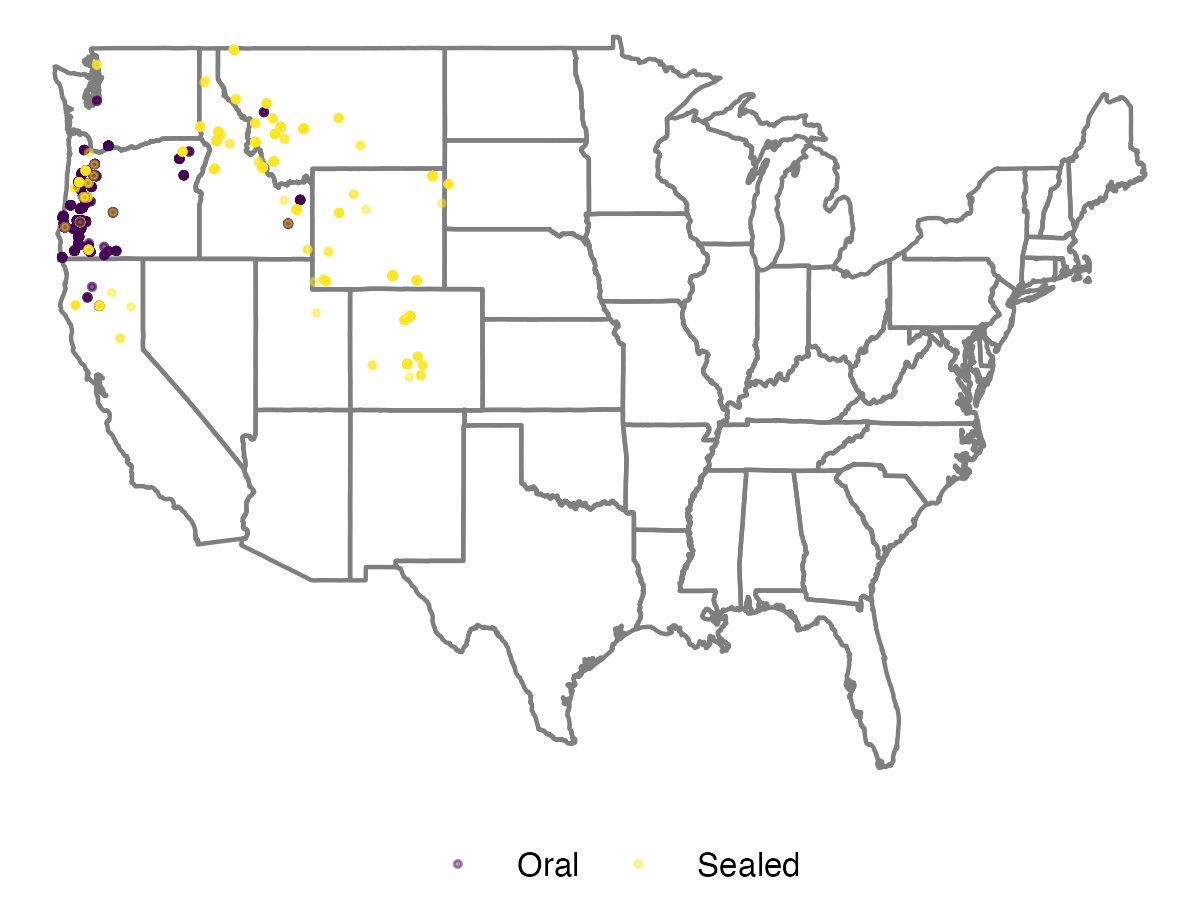}
\end{center}
\caption{Bidder address}\footnotesize
\label{fg:plot_bidder_address} 
\textit{Note}: We use the oral and sealed bid auction data from 2012Q2 to 2023Q1 by the Bureau of Land Management (BLM), U.S. Department of the Interior.
\end{figure}

Table \ref{tb:summary_statistics_of_entry_and_bidding_stage} reports summary statistics for both oral and sealed-bid auctions. Sealed bids are systematically smaller—in acres and estimated volumes—and close at lower dollar amounts, so they contribute less to aggregate revenue. Table \ref{tb:counterfactual_district_year_oral_auction_num} and Figure \ref{fg:plot_bidder_address} also show that oral auctions are more geographically concentrated, which lets us abstract from distance-related frictions such as transport costs. We therefore focus on oral auctions in the main text and relegate sealed-bid results to the appendix.

\begin{table}[ht!]\footnotesize
\begin{centering}
      \caption{Summary Statistics: Oral and Sealed Bid Auction}      \label{tb:summary_statistics_of_entry_and_bidding_stage} 
      \subfloat[Entry and bidding stage]{\input{figuretable/summary_statistics_of_entry_and_bidding_stage}}\\ 
      \subfloat[Cutting stage]{\input{figuretable/summary_statistics_of_cutting_stage}}\\
\end{centering}
 \footnotesize
  \textit{Note}: We use the oral and sealed bid auction data from 2012Q2 to 2023Q1 by the Bureau of Land Management (BLM), U.S. Department of the Interior.
\end{table} 

\begin{table}[!htbp]
  \begin{center}\footnotesize
      \caption{The distribution of district-year auctions }
      \label{tb:counterfactual_district_year_oral_auction_num} 
      \subfloat[Oral]{\input{figuretable/counterfactual_district_year_oral_auction_num}}\\
      \subfloat[Sealed]{\input{figuretable/counterfactual_district_year_sealed_auction_num}}
  \end{center}\footnotesize
  \textit{Note}: We use the oral and sealed bid auction data from 2012Q2 to 2023Q1 by the Bureau of Land Management (BLM), U.S. Department of the Interior.
\end{table}

\end{document}

%% file: figuretable/summary_statistics_of_entry_and_bidding_stage_oral.tex
\begin{tabular}[t]{lrrrrr}
\toprule
  & N & mean & sd & min & max\\
\midrule
Format &  & Oral &  &  & \\
Num of potential bidders & 306 & 10.34 & 8.75 & 1.00 & 29.00\\
Num of active bidders & 306 & 2.16 & 1.11 & 1.00 & 6.00\\
Manufacturer & 306 & 0.62 & 0.49 & 0.00 & 1.00\\
Entry rate & 306 & 0.43 & 0.36 & 0.03 & 1.00\\
Reserve price (\$1000) & 306 & 587.88 & 744.78 & 3.37 & 6060.53\\
Winning bid: $B_{1}$ (\$1000) & 306 & 754.23 & 934.18 & 3.37 & 6732.56\\
Second highest bid: $B_{2}$ (\$1000) & 209 & 843.76 & 985.58 & 10.18 & 6285.38\\
Bid ratio: $B_{2}/B_{1}$ & 209 & 0.98 & 0.05 & 0.66 & 1.00\\
Lumpsum & 306 & 0.86 & 0.35 & 0.00 & 1.00\\
Estimated volume (acres) & 306 & 228.29 & 185.47 & 0.00 & 2117.00\\
Estimated volume (mbf) & 304 & 3976.89 & 2945.76 & 8.50 & 16492.00\\
Estimated volume (ccf) & 304 & 6773.40 & 5007.61 & 13.90 & 26882.00\\
\bottomrule
\end{tabular}

%% file: figuretable/summary_statistics_of_cutting_stage_oral.tex
\begin{tabular}[t]{lrrrrr}
\toprule
  & N & mean & sd & min & max\\
\midrule
Format &  & Oral &  &  & \\
Num of cutting activities & 306 & 14.59 & 10.48 & 1.00 & 57.00\\
Cutting delay (avg month) & 306 & 21.99 & 13.88 & 0.97 & 131.71\\
Contract term (month) & 306 & 32.43 & 7.84 & 2.00 & 36.00\\
Revised contract term (month) & 306 & 36.72 & 13.41 & 1.97 & 111.29\\
Overdelayed dummy & 306 & 0.28 & 0.45 & 0.00 & 1.00\\
Total quantity (acres) & 306 & 194.83 & 165.42 & 0.00 & 1225.00\\
Total quantity (mbf) & 306 & 3672.37 & 3045.50 & 0.00 & 16501.50\\
Total quantity (ccf) & 306 & 5984.86 & 4965.41 & 0.00 & 26897.43\\
\bottomrule
\end{tabular}

%% file: figuretable/summary_statistics_of_diff_actualestimated.tex
\begin{tabular}[t]{lrrrrr}
\toprule
  & N & mean & sd & min & max\\
\midrule
Format &  & Oral &  &  & \\
Actual minus Estimated (acres) & 304 & -0.11 & 0.36 & -1.00 & 3.00\\
Actual minus Estimated (ccf) & 304 & -0.03 & 0.67 & -1.00 & 8.34\\
Actual minus Estimated (mbf) & 304 & -0.02 & 0.47 & -1.00 & 5.50\\
Actual minus Estimated (value) & 306 & -0.03 & 0.40 & -1.00 & 3.86\\
\bottomrule
\end{tabular}

%% file: figuretable/regression_result_transaction_cut_amount.tex
\begin{tabular}[t]{lccc}
\toprule
  & Transaction mbf & Transaction Acres & Transaction Value\\
\midrule
Intercept & -203.624*** & -2.300 & -56131.433***\\
 & (33.218) & (2.299) & (8155.076)\\
Lumber price & 1.334*** & 0.008 & 294.102***\\
 & (0.159) & (0.010) & (39.080)\\
Manufacturer & 57.458*** & -0.050 & 11256.345***\\
 & (13.317) & (0.875) & (3220.565)\\
Total Mbf & 0.032*** &  & \\
 & (0.002) &  & \\
Total Acres &  & 0.062*** & \\
 &  & (0.003) & \\
Total Value &  &  & 0.042***\\
 &  &  & (0.002)\\
\midrule
Num.Obs. & 4357 & 4357 & 4357\\
R2 & 0.104 & 0.100 & 0.210\\
R2 Adj. & 0.103 & 0.099 & 0.210\\
RMSE & 397.28 & 26.81 & 95662.45\\
\bottomrule
\end{tabular}

%% file: figuretable/poisson_result_num_of_active_bidders.tex
\begin{tabular}[t]{lc}
\toprule
  & Num of active bidders\\
\midrule
Intercept & 1.269*\\
 & (0.495)\\
Contract term & -0.003\\
 & (0.007)\\
Lumber price & -0.001\\
 & (0.002)\\
Covid dummy & -0.136\\
 & (0.313)\\
Reserve price & 0.000\\
 & \vphantom{1} (0.000)\\
Acres & 0.000\\
 & (0.000)\\
\midrule
Num.Obs. & 204\\
RMSE & 0.91\\
\bottomrule
\end{tabular}

%% file: figuretable/regression_result_share_of_active_bidder.tex
\begin{tabular}[t]{lc}
\toprule
  & \% of active bidders\\
\midrule
Intercept & 0.807**\\
 & (0.292)\\
Contract term & 0.000\\
 & (0.004)\\
Lumber price & -0.002\\
 & (0.001)\\
Covid dummy & -0.044\\
 & (0.180)\\
Reserve price & 0.000**\\
 & \vphantom{1} (0.000)\\
Acres & 0.000\\
 & (0.000)\\
\midrule
Num.Obs. & 204\\
RMSE & 0.36\\
\bottomrule
\end{tabular}

%% file: figuretable/regression_result_log_highest_bid_reserve_price.tex
\begin{tabular}[t]{lcccc}
\toprule
  & log(Winning price) & log(Winning price)  & log(Reserve price) & log(Reserve price) \\
\midrule
Intercept & 8.482*** & 8.471*** & 5.383*** & 6.861***\\
 & (0.574) & (0.543) & (0.750) & (0.632)\\
Contract term & 0.050*** & 0.046*** & 0.085*** & 0.061***\\
 & (0.008) & (0.008) & (0.011) & (0.009)\\
Lumber price & 0.008*** & 0.008*** & 0.019*** & 0.014***\\
 & (0.002) & (0.002) & (0.003) & (0.002)\\
Covid dummy & -0.219 & -0.378 & -0.979* & -0.998**\\
 & (0.339) & (0.321) & (0.467) & (0.383)\\
Num of active bidders & 0.171** & 0.162** & 0.058 & 0.033\\
 & (0.052) & (0.049) & (0.073) & (0.060)\\
Acres & 0.001** &  & 0.003*** & \\
 & (0.000) &  & (0.000) & \\
Mbf &  & 0.000*** &  & 0.000***\\
 &  & (0.000) &  & (0.000)\\
Manufacturer & 0.484*** & 0.474*** &  & \\
 & (0.108) & (0.102) &  & \\
\midrule
Num.Obs. & 204 & 204 & 204 & 204\\
R2 & 0.696 & 0.725 & 0.476 & 0.643\\
R2 Adj. & 0.685 & 0.716 & 0.463 & 0.633\\
RMSE & 0.66 & 0.63 & 0.93 & 0.77\\
\bottomrule
\end{tabular}

%% file: figuretable/auction_param_bias_rmse.tex
\begin{tabular}[t]{lrr}
\toprule
  & Bias & RMSE\\
\midrule
$\mu_l$ & 0.049 & 0.392\\
$\sigma_l$ & 0.005 & 0.153\\
$\mu_s$ & 0.209 & 0.994\\
$\sigma_s$ & -0.066 & 0.699\\
$\lambda_l$ & 0.001 & 0.028\\
$\lambda_s$ & -0.001 & 0.030\\
Sample size (Auction num) &  & 500\\
\bottomrule
\end{tabular}

%% file: figuretable/montecarlo_dynamic_bias_rmse.tex
\begin{tabular}[t]{lrr}
\toprule
  & Bias & RMSE\\
\midrule
$\gamma$ & -0.047 & 0.094\\
$c_{1}$ & -0.033 & 0.056\\
$c_{2}$ & -0.002 & 0.005\\
Sample size (Agent num) &  & 1000\\
\bottomrule
\end{tabular}

%% file: figuretable/cutting_stage_estimate.tex
\begin{tabular}[t]{lccc}
\toprule
 & $\gamma_{l}$ & $c_{1l}$ & $c_{2l}$\\
Estimates & 0.140 & 25.405 & 0.005\\
SE & {}[0.007] & {}[1.187] & {}[0.009]\\
 &  &  & \\
 & $\gamma_{s}$ & $c_{1s}$ & $c_{2s}$\\
Estimates & 0.038 & 7.242 & 0.000\\
SE & {}[0.026] & {}[4.768] & {}[0.009]\\
\bottomrule
\end{tabular}

%% file: figuretable/bidding_stage_estimate.tex
\begin{tabular}[t]{lcc}
\toprule
 & $\mu_{l}$ & $\sigma_{l}$\\
 & 0.988 & 0.768\\
 & {}[0.138] & {}[0.087]\\
 &  & \\
 & $\mu_{s}$ & $\sigma_{s}$\\
 & 8.166 & 1.065\\
 & {}[2.642] & {}[0.932]\\
\bottomrule
\end{tabular}

%% file: figuretable/entry_stage_estimate_oral.tex
\begin{tabular}[t]{lcc}
\toprule
 & $\lambda_{l}$ & $\lambda_{s}$\\
 & 0.161 & 0.153\\
 & {}[0.067] & {}[0.067]\\
\bottomrule
\end{tabular}

%% file: figuretable/counterfactual_winning_price_symmetric_small_tract_oral.tex
\begin{tabular}[t]{cccrrrr}
\toprule
Tract size & Format & Participants & 4 quarters & 8 quarters & 12 quarters & 16 quarters\\
\midrule
Small & Oral & (S, S) & 55.948 & 86.390 & 103.167 & 112.779\\
Small & Oral & (L, L) & 27.572 & 29.593 & 30.585 & 31.127\\
Small & Oral & (S, S, S) & 60.152 & 91.703 & 109.076 & 118.946\\
Small & Oral & (L, L, L) & 30.112 & 32.249 & 33.285 & 33.848\\
\bottomrule
\end{tabular}

%% file: figuretable/counterfactual_winning_price_symmetric_large_tract_oral.tex
\begin{tabular}[t]{cccrrrr}
\toprule
Tract size & Format & Participants & 4 quarters & 8 quarters & 12 quarters & 16 quarters\\
\midrule
Large & Oral & (S, S) & 176.821 & 322.606 & 413.583 & 470.140\\
Large & Oral & (L, L) & 152.029 & 163.126 & 169.948 & 174.185\\
Large & Oral & (S, S, S) & 184.277 & 332.675 & 425.201 & 482.485\\
Large & Oral & (L, L, L) & 158.072 & 169.357 & 176.308 & 180.624\\
\bottomrule
\end{tabular}

%% file: figuretable/counterfactual_winning_price_asymmetric_small_tract_oral.tex
\begin{tabular}[t]{cccrrrr}
\toprule
Tract size & Format & Participants & 4 quarters & 8 quarters & 12 quarters & 16 quarters\\
\midrule
Small & Oral & (S, L) & 30.176 & 32.322 & 33.352 & 33.887\\
Small & Oral & (S, S, L) & 55.949 & 86.390 & 103.167 & 112.779\\
Small & Oral & (S, L, L) & 32.847 & 35.107 & 36.187 & 36.741\\
\bottomrule
\end{tabular}

%% file: figuretable/counterfactual_winning_price_asymmetric_large_tract_oral.tex
\begin{tabular}[t]{cccrrrr}
\toprule
Tract size & Format & Participants & 4 quarters & 8 quarters & 12 quarters & 16 quarters\\
\midrule
Large & Oral & (S, L) & 157.572 & 169.365 & 176.312 & 180.62\\
Large & Oral & (S, S, L) & 177.604 & 322.606 & 413.583 & 470.14\\
Large & Oral & (S, L, L) & 163.453 & 175.716 & 182.794 & 187.19\\
\bottomrule
\end{tabular}

%% file: figuretable/summary_statistics_of_entry_and_bidding_stage.tex
\begin{tabular}[t]{lrrrrrrrrr}
\toprule
  & N & mean & sd & min & max & mean & sd & min & max\\
\midrule
Format &  & Oral &  &  &  & Sealed &  &  & \\
Num of auctions &  & 306 &  &  &  & 112 &  &  & \\
Num of potential bidders & 418 & 10.34 & 8.75 & 1.00 & 29.00 & 5.25 & 7.36 & 1.00 & 29.00\\
Num of active bidders & 418 & 2.16 & 1.11 & 1.00 & 6.00 & 2.03 & 1.20 & 1.00 & 6.00\\
Manufacturer & 418 & 0.62 & 0.49 & 0.00 & 1.00 & 0.62 & 0.49 & 0.00 & 1.00\\
Entry rate & 418 & 0.43 & 0.36 & 0.03 & 1.00 & 0.70 & 0.34 & 0.03 & 1.00\\
Reserve price (\$1000) & 418 & 587.88 & 744.78 & 3.37 & 6060.53 & 260.35 & 546.12 & 0.58 & 3427.69\\
Winning bid: $B_{1}$ (\$1000) & 418 & 754.23 & 934.18 & 3.37 & 6732.56 & 380.99 & 811.36 & 1.05 & 5193.26\\
Second highest bid: $B_{2}$ (\$1000) & 270 & 843.76 & 985.58 & 10.18 & 6285.38 & 439.75 & 789.56 & 0.96 & 4061.53\\
Bid ratio: $B_{2}/B_{1}$ & 270 & 0.98 & 0.05 & 0.66 & 1.00 & 0.74 & 0.19 & 0.15 & 1.00\\
Lumpsum & 418 & 0.86 & 0.35 & 0.00 & 1.00 & 0.47 & 0.50 & 0.00 & 1.00\\
Estimated volume (acres) & 418 & 228.29 & 185.47 & 0.00 & 2117.00 & 176.62 & 179.19 & 2.00 & 989.00\\
Estimated volume (mbf) & 415 & 3976.89 & 2945.76 & 8.50 & 16492.00 & 2080.19 & 2587.46 & 30.00 & 13374.00\\
Estimated volume (ccf) & 415 & 6773.40 & 5007.61 & 13.90 & 26882.00 & 3564.90 & 4563.34 & 48.90 & 23077.60\\
\bottomrule
\end{tabular}

%% file: figuretable/summary_statistics_of_cutting_stage.tex
\begin{tabular}[t]{lrrrrrrrrr}
\toprule
  & N & mean & sd & min & max & mean & sd & min & max\\
\midrule
Format &  & Oral &  &  &  & Sealed &  &  & \\
Num of auctions &  & 306 &  &  &  & 112 &  &  & \\
Num of cutting activities & 418 & 14.59 & 10.48 & 1.00 & 57.00 & 5.52 & 5.76 & 1.00 & 27.00\\
Cutting delay (avg month) & 418 & 21.99 & 13.88 & 0.97 & 131.71 & 15.62 & 12.88 & 1.59 & 72.00\\
Contract term (month) & 418 & 32.43 & 7.84 & 2.00 & 36.00 & 26.55 & 10.01 & 2.00 & 36.00\\
Revised contract term (month) & 418 & 36.72 & 13.41 & 1.97 & 111.29 & 30.50 & 13.66 & 1.90 & 71.55\\
Overdelayed dummy & 418 & 0.28 & 0.45 & 0.00 & 1.00 & 0.09 & 0.29 & 0.00 & 1.00\\
Total quantity (acres) & 418 & 194.83 & 165.42 & 0.00 & 1225.00 & 161.31 & 198.45 & 0.00 & 1157.00\\
Total quantity (mbf) & 418 & 3672.37 & 3045.50 & 0.00 & 16501.50 & 1921.33 & 3081.59 & 0.00 & 25231.30\\
Total quantity (ccf) & 418 & 5984.86 & 4965.41 & 0.00 & 26897.43 & 3106.07 & 5027.84 & 0.00 & 41127.04\\
\bottomrule
\end{tabular}

%% file: figuretable/counterfactual_district_year_oral_auction_num.tex
\begin{tabular}[t]{cccccccccccc}
\toprule
District & 2012 & 2013 & 2014 & 2015 & 2016 & 2017 & 2018 & 2019 & 2020 & 2021 & 2022\\
\midrule
Coos Bay DO & 19 & 5 & 5 & 5 & 0 & 2 & 2 & 2 & 1 & 0 & 2\\
Eugene DO & 8 & 6 & 5 & 3 & 3 & 0 & 0 & 0 & 0 & 0 & 0\\
Medford DO & 7 & 3 & 1 & 5 & 0 & 1 & 0 & 0 & 1 & 1 & 0\\
Roseburg DO & 19 & 7 & 4 & 6 & 3 & 2 & 1 & 2 & 2 & 0 & 5\\
Salem DO & 10 & 3 & 9 & 3 & 3 & 0 & 0 & 0 & 0 & 0 & 0\\
Lakeview DO & 0 & 1 & 0 & 1 & 1 & 1 & 0 & 0 & 0 & 0 & 0\\
Northern California DO & 0 & 1 & 0 & 0 & 0 & 1 & 0 & 0 & 0 & 0 & 0\\
Idaho Falls DO & 0 & 0 & 1 & 1 & 0 & 1 & 0 & 0 & 0 & 0 & 0\\
Roseburg & 0 & 0 & 0 & 1 & 0 & 0 & 0 & 0 & 0 & 0 & 0\\
NW Oregon DO & 0 & 0 & 0 & 0 & 0 & 4 & 10 & 4 & 8 & 2 & 4\\
\bottomrule
\end{tabular}

%% file: figuretable/counterfactual_district_year_sealed_auction_num.tex
\begin{tabular}[t]{cccccccccccc}
\toprule
District & 2012 & 2013 & 2014 & 2015 & 2016 & 2017 & 2018 & 2019 & 2020 & 2021 & 2022\\
\midrule
Idaho Falls DO & 1 & 1 & 0 & 0 & 0 & 0 & 2 & 1 & 0 & 0 & 3\\
Western Montana DO & 2 & 0 & 1 & 0 & 4 & 1 & 2 & 1 & 1 & 0 & 0\\
Wyoming High Desert DO & 0 & 1 & 0 & 1 & 0 & 0 & 0 & 0 & 0 & 0 & 0\\
Coeur d'Alene DO & 0 & 0 & 1 & 0 & 1 & 0 & 1 & 0 & 1 & 0 & 1\\
Front Range DO & 0 & 0 & 2 & 1 & 0 & 0 & 0 & 0 & 0 & 0 & 0\\
Salem DO & 0 & 0 & 1 & 0 & 0 & 0 & 0 & 0 & 0 & 0 & 0\\
Northwest DO & 0 & 0 & 0 & 1 & 0 & 0 & 1 & 0 & 0 & 0 & 0\\
Central Montana DO & 0 & 0 & 0 & 0 & 0 & 0 & 1 & 0 & 0 & 0 & 0\\
Spokane DO & 0 & 0 & 0 & 0 & 0 & 0 & 1 & 0 & 0 & 0 & 2\\
Boise DO & 0 & 0 & 0 & 0 & 0 & 0 & 0 & 1 & 1 & 0 & 0\\
Wyoming High Plains DO & 0 & 0 & 0 & 0 & 0 & 0 & 0 & 0 & 1 & 0 & 0\\
Coos Bay DO & 0 & 0 & 0 & 0 & 0 & 0 & 0 & 0 & 0 & 2 & 0\\
NW Oregon DO & 0 & 0 & 0 & 0 & 0 & 0 & 0 & 0 & 0 & 4 & 6\\
North Central Montana District & 0 & 0 & 0 & 0 & 0 & 0 & 0 & 0 & 0 & 1 & 0\\
Northern California DO & 0 & 0 & 0 & 0 & 0 & 0 & 0 & 0 & 0 & 1 & 0\\
Roseburg DO & 0 & 0 & 0 & 0 & 0 & 0 & 0 & 0 & 0 & 4 & 2\\
West Desert DO & 0 & 0 & 0 & 0 & 0 & 0 & 0 & 0 & 0 & 0 & 1\\
\bottomrule
\end{tabular}